\documentclass[aps,prd,superscriptaddress,showpacs,preprint,nofootinbib]{revtex4}
\usepackage{graphicx, bm}
\usepackage{amsmath}
\usepackage{mathrsfs}
\usepackage{amsmath}
\usepackage{amssymb}
\usepackage{bbold}
\usepackage[usenames]{color}
\usepackage{float}
\usepackage{subfigure}

\usepackage{graphicx,color}
\usepackage{amsmath,amssymb}
\usepackage{url}
\usepackage{epstopdf}

\newcommand{\hs}{\hspace*{0.5cm}}

\newcommand{\be}{\begin{equation}}
\newcommand{\ee}{\end{equation}}
\newcommand{\bea}{\begin{eqnarray}}
\newcommand{\eea}{\end{eqnarray}}
\newcommand{\nn}{\nonumber}
\newcommand{\crn}{\nonumber \\}

\newcommand{\al}{\alpha}
\newcommand{\la}{\lambda}
\newcommand{\bet}{\beta}
\newcommand{\ga}{\gamma}

\newcommand{\om}{\omega}

\newcommand{\fr}{\frac}
\newcommand{\bc}{\begin{center}}
	\newcommand{\ec}{\end{center}}
\newcommand{\Ga}{\Gamma}
\newcommand{\de}{\delta}

\newcommand{\ep}{\epsilon}

\newcommand{\si}{\sigma}

\newcommand {\ba}{\begin{array}}
\newcommand {\ea}{\end{array}}
\newcommand{\ben}{\begin{enumerate}}
\newcommand{\een}{\end{enumerate}}

\begin{document}
%\tightenlines

\title{Neutrino Energy Loss Rates  in 3-3-1 Models }

\author{D. T. Binh \footnote{dinhthanhbinh3@duytan.edu.vn}}
\affiliation{\small Institute for Theoretical and Applied Research\\
	Faculty of Natural Science, Duy Tan University\\}

\author{L. T. Hue \footnote{lthue1981@gmail.com}}
\affiliation{Institute of Physics, Vietnam  Academy of Science and Technology, 10 Dao Tan, Ba Dinh, Hanoi, Vietnam}

\author{ V. H. Binh \footnote{vhbinh@iop.vast.ac.vn}}
\affiliation{Institute of Physics, Vietnam  Academy of Science and Technology, 10 Dao Tan, Ba Dinh, Hanoi, Vietnam}
\affiliation{ Graduate University of Science and Technology,
	Vietnam Academy of Science and Technology,
	18 Hoang Quoc Viet, Cau Giay, Hanoi, Vietnam}	
\author{ H. N. Long \footnote{hnlong@iop.vast.ac.vn}}

\affiliation{Institute of Physics, Vietnam  Academy of Science and Technology, 10 Dao Tan, Ba Dinh, Hanoi, Vietnam}
\date{\today}

\begin{abstract}

The stellar energy-loss rates $\mathcal{Q}$  due to the production of neutrino pair in the framework of  3-3-1 models are presented. The energy loss rate $\mathcal{Q}$ is evaluated for different values of  $\beta=\pm\fr{1}{\sqrt{3}},\pm\fr{2}{\sqrt{3}},\pm\sqrt{3}$ in which $\beta$ is a parameter used to define the charge operator in the 3-3-1 models. The correction to the rate which is compared with that of the Standard Model ($\de \mathcal{Q}$) is also evaluated. We show that the correction does not exceed 14\% and %is
gets the highest with $\beta=-\sqrt{3}$.
The contribution of dipole moment to the energy loss rate is small compared to the contribution of new natural gauge boson $Z'$ and this  sets constraints for the mass of Z' $m_{Z'} \leq 4000$ GeV. This mass range is within the searching range for $Z'$ boson at LHC.
\end{abstract}


\vspace{5mm}

\maketitle

\section{Introduction}

Cooling rate is an important factor that affects the stellar evolution. %During their life time,
In their life time,
stars can emit energy in the form of radiation, flux of neutrinos \cite{Particle-Astro,Axion-Eloss-Raffelt,neutrino-Eloss-Beaudet,neutrino-Eloss-esposito2002,neutrino-Eloss-esposito2003,Neutrino-Gamow-1,Neutrino-Gamow-2,Neutrino-Pontecorvo},  gravitational wave \cite{Gwave-emission-1,Gwave-emission-2} and  axion \cite{Axion-Flux-1,Axion-Flux-2,Axion-Flux-3}. It is well known that  cooling by neutrino emission plays an important role in
a variety of stellar systems, from  neutron stars and core-collapse supernovae (CCSNe), from  low-mass red giants (RG)
and horizontal branch (HB) stars to white dwarfs (WDs). Despite of the fact that
neutrinos interact extremely weakly,
once produced they easily escape from stellar interiors carrying
away energy.
The neutrino production is main process (99\%) while photoproduction is tiny one (1-2 \%) in a massive star collapse.

There are mainly four interaction mechanisms for the energy loss due to neutrino emissions: (i)  pair annihilation \hs $e^+ + e^- \rightarrow \nu + \bar{ \nu}$; (ii) $\nu$ photoproduction \hs  $\gamma + e^{\pm} \rightarrow e^{\pm}  + \nu +\bar{ \nu}$ ; (iii) plasmon decay \hs   $\gamma^* \rightarrow \nu + \bar{ \nu}$; and (iv)  Bremsstrahlung on nuclei   \hs  $e^{\pm}+Z \rightarrow e^{\pm} +Z + \nu +\bar{ \nu}$.
% mechanism:
Each of these processes  will give a  dominant contribution to the energy loss rate ($\mathcal{Q}$) in a particular  region of temperature and density ( corresponding to a certain evolution period of the star. The pair annihilation process dominates in a high temperature ($T \geq 10^9$ K) and not too high density ($\rho$). The  $\nu $ photoproduction on the other hand give leading contribution in regions where $  10^{8 } \textrm{K} \leq T\leq 10^{10}$ K and low densities ($\rho \leq 10^5\; \mathrm{gcm^{-3}}$). Finally, plasmon decay and bremsstrahlung on nuclei are dominants process for large ($\rho \geq 10^6\; \mathrm{gcm^{-3}}$) and very large ($\rho \geq $$10^9 gcm^{-3}$) core densities respectively with temperature in range  $  10^{8 } \textrm{K} \leq T\leq 10^{10} $K.

 In the Standard Model (SM) frame work \cite{SM-Glashow,SM-Weinberg,SM-Salam}, the energy loss rate (ELR) due to neutrino emission ($\mathcal{Q}^\nu$) receives contributions from both weak nuclear reactions and purely leptonic processes. However, in many models beyond the SM (BSM)  new interactions or new contributions from new particles can change the rate at which neutrinos are produced therefore the evolution of star may be modified.   The stellar energy loss rate was calculated in the frame works of the SM \cite{QSM-Dicus1972} and the $U(1)_{B-L}$ extension model of SM~\cite{QBSM-Hernandez}.

In the SM,  neutrinos are massless therefore neutrinos photon interaction at tree level do not exits. However, neutrinos oscillation experiment \cite{SNO,ATM}  imply that neutrinos do have mass. In some beyond  SM models,
neutrino can be massive. Consequently,
 there exist dipole moments. The interaction of neutrino with photon  via dipole moment can affect the ELR \cite{Dipole-Eloss-Heger2009,Dipole-Eloss-Kerimov1992,QBSM-Hernandez}.

New natural gauge boson $Z'$ appears naturally in some extension models of the SM such as the  Left-Right symmetric model \cite{G.Senjanovic,G.Senjanovic1},  the model of composite boson \cite{Baur}. One of the simplest and attractive extension of the SM is the  $SU(3)_L$ extension of the SM \cite{ppf-1,ppf-2,ppf-3,r331-1,r331-2,r331-3,r331-4,r331-5,331beta1,331beta2}, %Diaz:2004fs},
where the SM  fermion doublets are assigned  as $SU(3)_L$ triplets or antitriplets including new exotic   fermions or positrons in the third components of the $SU(3)_L$ (anti) triplets. In this work we pay attention to the 3-3-1 model with an arbitrary parameter $\beta$  (3-3-1$\beta$)  containing  exotic fermions with electric charges defined  by the charge operator characterized by  $\beta$.  In general the class of 331 models have the same characteristics as follows:
1) The anomaly in 3-3-1 model is canceled when all fermion generations are considered, 2) Peccei-Quinn (PQ) symmetry~\cite{Peccei-Quin} is a
 result of gauge invariance in the model 3) As the extension of the gauge group there appears new natural gauge boson $Z'$, 4) One generation of quark is different from the other two ones,  leading to the appearance of the  tree level Flavor Changing Neutral Current (FCNC) through the mixing $Z-Z'$.  Also, the interactions of the $Z'$ and neutrinos  will affect the production rates of neutrinos and modify the energy loss rate  predicted by the SM.
%The $Z'$  boson can interaction with neutrinos in the form of V-A interaction therefore modifies the producing rate of neutrinos.

There are may works on the stellar energy loss in the frame work of the SM \cite{QSM-Dicus1972} and extension models of
the SM \cite{QBSM-Hernandez,QBSM-Bugarin}. In this work we will investigate the effect of magnetic dipole moment and  new $Z'$ boson  on the % energy loss rate
ELR of a stellar. In our work, we will investigate the
% energy loss rate
ELR  through the process  $e^+ e^- \rightarrow (\ga, W, Z, Z') \rightarrow \nu \bar{ \nu}$. We investigate the energy loss rate of the $331\beta $ model and its relative correction compared with the SM.

Our work is organized as follows: in section II we will briefly  review  the $331\beta$. In section III we will calculate the $e^+e^-\rightarrow \nu\overline{\nu}$  amplitude, derive  its analytical approximation in different limits. Lastly, section IV and V are the numerical discussion and conclusion.

\section{ The model 3-3-1$\beta$ }
\label{models}

  The model 3-3-1$\beta$ is constructed based on the gauge group $SU(3)_c\times SU(3)_L\times U(1)_X$. One common feature of the class  of $SU(3)_L$ model is that the extension of the gauge group from $SU(2)_L\rightarrow SU(3)_L$ requires new fermions. Normaly, the left-handed fermions are arranged into the third components of the triplets, while the right ones are in the $SU(3)_L$ singlets.  The anomaly cancellation requires that the number of fermion triplets equals the number of fermion antitriplets,  leading to the consequence that one quark family must have the same $SU(3)_L$ representation as the three lepton families and different from the remaining quark families. The electric charges of all particles in the 3-3-1$\beta$  are determined by the following charge operator

\be
Q = I_3 + \bet \,  I_8 + X\, ,
\label{eq:charge_Q}
\ee
where $I_3$, $I_8$ are  the $SU(3)$ generators.
The models are characterized by the parameter  $\beta$ in  the charge operator $Q$.
% depends on two  parameters $\beta$ and $X$.
The lepton representation can be represented as follows \cite{331beta1,331beta2}:
\bea && L'_{aL}=\left(
\begin{array}{c}
	e'_a \\
	-\nu'_{a} \\
	E'_a \\
\end{array}
\right)_L \sim \left(1,3^*~, -\fr{1}{2}+\fr{\beta}{2\sqrt{3}}\right), \hs a=1,2,3,\crn
&& e'_{aR}\sim   \left(1, 1~, -1\right)  , \hs \nu'_{aR}\sim  \left(1, ~1~, 0\right) ,\hs E'_{aR} \sim   \left(1, ~1~, -\fr{1}{2}+\fr{\sqrt{3}\beta}{2}\right).  \label{lep}
\eea

In particular, the left-handed leptons are assigned to anti-triplets while the right-handed leptons to  singlets.  The model predicts three  exotic leptons $E'^a_{L,R}$ which are much heavier than the ordinary leptons.  The right-handed neutrinos $\nu'_{aR}$ is needed to generate Dirac mass for active neutrinos
The prime denotes flavor states to be distinguished with mass eigenstates will be introduced later.
The numbers in the parentheses are to label the representation of $SU(3)_C \otimes SU(3)_L\otimes U(1)_X$ group.

For our purpose of this work, the quark sector is irrelevant therefore we do not present it here. It has been discussed in details in many previous works~\cite{331beta1,331beta2,331beta3, Buras:2012jb}.

The detail calculation of gauge and Higgs interactions
has been shown in \cite{331beta1,331beta2,331beta3,Buras:2012jb}.
Within nine  EW gauge bosons, the  covariant derivative is defined as follows
\be D_{\mu}\equiv \partial_{\mu}-i g I^a W^a_{\mu}-i g_X X I^9X_{\mu},  \label{coderivative1}
\ee
where $I^9=\fr{\mathbb{1}}{\sqrt{6}}$, $g$ and $g_X$ are coupling constants corresponding to the two groups $SU(3)_L$ and $U(1)_X$, respectively.
The matrix $W^a I^a$  for a triplet
 can be written as
\bea W^a_{\mu}I^a=\fr{1}{2}\left(
\begin{array}{ccc}
	W^3_{\mu}+\fr{1}{\sqrt{3}} W^8_{\mu}& \sqrt{2}W^+_{\mu} &  \sqrt{2}Y^{+A}_{\mu} \\
	\sqrt{2}W^-_{\mu} &  -W^3_{\mu}+\fr{1}{\sqrt{3}} W^8_{\mu} & \sqrt{2}V^{+B}_{\mu} \\
	\sqrt{2}Y^{-A}_{\mu}& \sqrt{2}V^{-B}_{\mu} &-\fr{2}{\sqrt{3}} W^8_{\mu}\\
\end{array}
\right),
\label{wata}\eea
where we have denoted
% the mass eigenstates of
the charged gauge bosons as
\bea W^{\pm}_{\mu}=\fr{1}{\sqrt{2}}\left( W^1_{\mu}\mp i W^2_{\mu}\right),\crn
Y^{\pm A}_{\mu}=\fr{1}{\sqrt{2}}\left( W^4_{\mu}\mp i W^5_{\mu}\right),\crn
V^{\pm B}_{\mu}=\fr{1}{\sqrt{2}}\left( W^6_{\mu}\mp i W^7_{\mu}\right).
\label{gbos}\eea
From (\ref{eq:charge_Q}), the electric charges of the gauge bosons are
given by
\bea
A=\fr{1}{2}+\beta\fr{\sqrt{3}}{2}, \hs
B=-\fr{1}{2}+\beta\fr{\sqrt{3}}{2}\label{charge_AB}.\eea

The scalar sector contains
 three scalar triplets as follows
 %They are defined as
\bea && \chi=\left(
\begin{array}{c}
	\chi^{+A} \\
	\chi^{+B} \\
	\chi^0 \\
\end{array}
\right)\sim \left(1, 3~, \fr{\beta}{\sqrt{3}}\right), \hs
\eta=\left(
\begin{array}{c}
	\eta^0 \\
	\eta^- \\
	\eta^{-A} \\
\end{array}
\right)\sim \left(1, 3~, -\fr{1}{2}-\fr{\beta}{2\sqrt{3}}\right)
\crn
&&
 \rho=\left(
\begin{array}{c}
	\rho^+ \\
	\rho^0 \\
	\rho^{-B} \\
\end{array}
\right)\sim \left(1, 3~, \fr{1}{2}-\fr{\beta}{2\sqrt{3}}\right),
\label{higgsc}
\eea
where $A,B$ denote electric charges as
determined in (\ref{charge_AB}).  Only the vacuum expectation values (VEV) of the neutral Higgs components are non zero and defined as follows: $\langle  \chi^0\rangle=\om/\sqrt{2}$, $\langle  \rho^0\rangle=v/\sqrt{2}$,  and $\langle  \eta^0\rangle=u/\sqrt{2}$. They are enoungh to generate masses for all particles in the model.

As usual, the symmetry breaking  happens in two steps:
$SU(3)_L\otimes U(1)_X\xrightarrow{ \om} SU(2)_L\otimes U(1)_Y\xrightarrow{v,u} U(1)_Q$.
Therefore, it is reasonable to assume that $\om \gg  v,u$. There are well-known relations between the gauge couplings of the 3-3-1$\beta$ model and the SM, namely
\be
g_2 = g,\hs  \fr{g_X^2}{g^2} = \fr{6s_W^2}{1-(1+\beta^2)s_W^2},,
\label{matching_coupl}
\ee
where $g_2$ and $g_1$ are the  couplings
 corresponding to $SU(2)_L$ and $U(1)_Y$ subgroups,  respectively. The weak mixing angle is defined as   $ \sin\theta_W \equiv s_W$,  $\tan\theta_W \equiv t_W =  \fr{g_1}{g_2}$,
 	and so forth.

The equation in (\ref{matching_coupl}) leads to an interesting constraint of the parameter $\beta$:
\bea
|\beta| \le \sqrt{3}\,.
\label{eq_beta_constraint}
\eea

With the above VEVs,  the charged gauge boson masses are
\bea
m^2_{Y^{\pm A}} = \fr{g^2}{4}(\om^2+u^{ 2}),\hs
m^2_{V^{\pm B}}=\fr{g^2}{4}(\om^2+v^2),\hs
m^2_W = \fr{g^2}{4}(v^2 + u^{ 2})\, .
\label{masga}\eea

Let us now discuss the mixings of leptons. In the 3-3-1$\beta$ with heavy exotic leptons, we will ignore the mixing betweent the SM leptons and these new leptons  in the case  that they have the same electric charges.
 The Yukawa
 interations related to the above mentioned  mixings
 given by
\bea -\mathcal{L}^\text{Yuk}_{\mathrm{lepton}}= Y^e_{ab} \overline{L'}_{aL} \eta^*e'_{bR}
+ Y^\nu_{ab} \overline{L'}_{aL} \rho^*\nu'_{bR} + Y^E_{ab} \overline{L'}_{aL} \chi^*E'_{bR}
+\mathrm{H.c.}, \label{ylepton1}\eea
where $a,b=e,\mu,\tau$ are family indices. This Largangian generates  consistent masses for leptons. Hereafter, without loss of generality, ones will work in the basis where the SM charged leptons are in their mass eigenstates.  Ones
	can therefore set $Y^e_{ab}$ to be diagonal and $e' = e$ in Eqs.~\eqref{ylepton1}.
 The SM lepton masses  are $m_{e_a}=Y^e_{aa}u/\sqrt{2}$.  The mass term and  mixing of the Dirac neutrinos are derived as follows:
\bea -\mathcal{L}^\text{mass}_{\nu}= \fr{Y^\nu_{ab}v}{\sqrt{2}} \overline{\nu'}_{aL} \nu'_{bR}
+\mathrm{H.c.},\hs  \nu'_{aL} = U^L_{ab} \nu_{bL}, \hs \nu'_{aR} = U^R_{ab} \nu_{bR}, \label{mlepton1}\eea
where $U^{L,R}$ are $3\times 3$ unitary matrices the neutrinos, respectively. It can be identified that   $U^L=U_\text{PMNS}$ are the well-known lepton mixing matrix.

The Higgs sector does not involve our work. The mass and eigenstates of the all Higgs in the 3-3-1$\beta$ model were given detailedly previously, for example ref.~\cite{331beta2}.  Hence, we will not repeat here. We stress that  the scalar sector contains six charged Higgs bosons, one neutral pseudoscalar Higgs and three neutral scalar Higgs bosons, which one of them can be identified with the standard model-like Higgs found by experiments at LHC.

The neutral currents mediated by $Z$ and $Z'$ bosons relating with the lepton sector used in our calculation are given by:

\be
i\,L_{int}^Z = \frac{ i g}{2 c_W} Z^\mu
\sum_{\ell=e,\mu,\tau} \left[ {\bar \nu}_{\ell \,L} \gamma_\mu \nu_{\ell L}-(1 -2 s_W^2) {\bar \ell}_L \gamma_\mu \ell_L + 2 s_W^2 {\bar \ell}_R \gamma_\mu \ell_R \right],
\ee
and
\begin{align}
\label{ZprimeFR}
i\,L_{int}^{Z^\prime} =&  i\frac{g  Z'^{\mu}}{ 2 \sqrt{3} c_W \sqrt{1-(1+\beta^2) s_W^2}} \nn\\
&\times   \sum_{\ell=e,\mu,\tau}\left\{\left[1-(1+\sqrt{3} \beta) s_W^2 \right] \left({\bar \nu}_{\ell \, L} \gamma_\mu \nu_{\ell \, L}+ {\bar \ell}_L \gamma_\mu \ell_L \right)- 2 \sqrt{3} \beta s_W^2 {\bar \ell}_R \gamma_\mu \ell_R\right\}.
\end{align}

The common $V-A$ form of 
 the interaction of the Z bosons with  $e, \nu_e$ are \cite{331beta3}

\be
\mathcal{L}_{Z^i ff}=
\fr{g}{2c_W}\bar{ f}\ga_\mu [g^{Z^i}_V(f)-g^{Z^i}_A(f)\ga _5]fZ_i^{\mu}
\ee
where $Z_i=Z, Z'$ and the $g^{Z^i}_V(f)$, $ g^{Z^i}_A(f)$ are given in Table \ref{Zff}.% and  \ref{Z'ff}, respectively.

The common $V-A$ form of the Lagrangian of  the interactions of the  neutral gauge  bosons with  $e, \nu_e$ are
 \be
\mathcal{L}_{Z^i ff}=\frac{g}{2c_W}\bar{ f}\ga^\mu [g^{Z^i}_V(f)-g^{Z^i}_A(f)\ga_5]fZ^i_\mu\,,
\ee
where $Z^i=Z, Z'$.  The $g^{Z^i}_V(f)$, $ g^{Z^i}_A(f)$ are given in Table \ref{Zff}.

\begin{table}[ht!]
	\centering
\resizebox{16cm}{!}{
	\begin{tabular}{|c|c|c|c|c|}
		\hline
		$f | Z^i$ & $g^Z_V(f)$ & $g^Z_A(f)$ &$g^{Z'}_V(f)$ &  $g^{Z'}_A(f)$\\
		\hline
		$\nu$ &  $\fr 1 2$ & $\fr 1 2$ &  $  \frac{1}{2\sqrt{3}} \frac{[1-(1+\sqrt{3}\beta)s_w^2]}{\sqrt{1-(1+\beta^2)s_w^2}}$ & $  \frac{1}{2\sqrt{3}} \frac{[1-(1+\sqrt{3}\beta)s_w^2]}{\sqrt{1-(1+\beta^2)s_w^2}}$\\
		\hline
		$e$ & $-\fr 1 2 + 2 s^2_W$ & $-\fr 1 2 $& $  \frac{1}{2\sqrt{3}} \frac{[1-(1+3\sqrt{3}\beta)s_w^2]}{\sqrt{1-(1+\beta^2)s_w^2}}$ & $  \frac{1}{2\sqrt{3}} \frac{[1-(1-\sqrt{3}\beta)s_w^2]}{\sqrt{1-(1+\beta^2)s_w^2}}$\\
		\hline
	\end{tabular}
}
	\caption{The couplings of $Z$ and $Z'$ with  $e, \nu_e$.}
	\label{Zff}
\end{table}	

%%%%%%%%%%%%%%

\section{Stellar ELRs
 through  annihilation of electron-positron pair into electron neutrino and antineutrino
}
\label{stellenergy}

Let us consider the process inside star, namely the annihilation of electron-positron pair into electron neutrino and antineutrino:

\be
e^{+}\left(p_1 \right) e^{-}\left(p_2 \right)
\rightarrow (\ga , W, Z_a)\to \nu_i\left(k_1,\la _3 \right)\bar\nu_i\left(k_2,\la _4 \right).
\label{process}
\ee

\begin{figure}[ht!]
	\centering
	\includegraphics[scale=1.0]{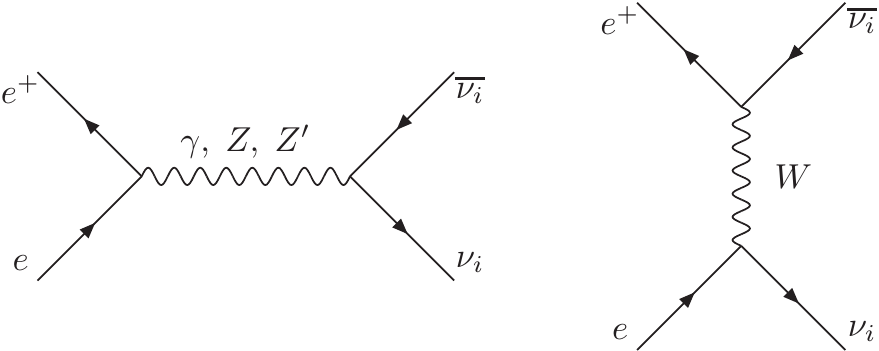}
	\caption{Feynman diagrams contributing  to $e^+ e^- \rightarrow \nu \bar{\nu}$}
	\label{FeyDiagrams}
\end{figure}

The Feynman diagrams contributing
to the process given by Eq. (\ref{process}) are shown in Fig.\ref{FeyDiagrams} where  $Z_a=Z, Z'$ and
 $\nu_i=\nu_e, \nu_\mu, \nu_\tau$, respectively. Here, $ p_1, p_2 $ are the momentum of the incoming electron, positron
and $ k_1,k_2 $ are the momentum of the outgoing $\nu \bar{ \nu}$ pair while $\la _{3,4}$ is the neutrino helicity.

In Fig.\ref{FeyDiagrams}, the coupling between neutrinos and photon %can
arises at the quantum
level from loop diagrams.
The general expression for the electromagnetic form factors of a massive neutrino is as follows \cite{Neutrino-FFactor-Kayser,Neutrino-FFactor-Nieves,Neutrino-FFactor-Broggini,EMproperties-Neutrino}

\be
\Ga ^{\mu}(q)=eF_{1}(q^{2})\ga ^{\mu}+\fr{ie}{2m_{\nu}}F_{2}(q^{2})\si ^{\mu \nu}q_{\nu}
+\fr{e}{2m_\nu}F_3(q^2)\ga _5\si ^{\mu \nu}q_\nu +eF_4(q^2)\ga _5\left(\ga ^\mu -\fr{q\llap{/}q^\mu}{q^{2}}\right),
\ee
where $q^\mu$ is the photon momentum, and $F_i(q^2), \, , \, i=1, 2, 3, 4$ are the electromagnetic
form factors of the neutrino. In this analysis, we are interested in the anomalous magnetic moment (AMM)
and the electric dipole moment (EDM) of the neutrino, which are defined in terms of the
$ F_2 $ and $ F_3 $ form factors at $ q^2=0$  as follows:

\be
\mu_\nu=\fr{m_e}{m_\nu}F_2|_{q^2=0},  \hs d_\nu=\fr{e}{2m_\nu} F_3|_{q^2=0}\, .
\ee

\subsection{Low energy  amplitudes}

For our purpose, the low energy limit where the  propagator of gauge bosons  takes a form $\fr{-ig_{\mu \nu}}{M^2}$ is applicable.
In the mentioned  limit, the
amplitudes are thus given by:
\bea
i\mathcal{M}^\ga _{fi}&=&\bar{u}\left(k_2,\la _4\right)\Ga ^{\mu}v\left(k_1,\la _3\right)
\fr{g_{\mu\nu}}{\left(p_1+p_2 \right)^2}
\bar{v}\left(p_1\right) e\ga ^\nu u\left(p_2\right),\\
i\mathcal{M}^{Z}_{fi}&=&- \left( \fr{g^2}{8\cos^2{\theta_W}M_{Z}^2}\right)
\bar{u}\left(k_2,\la _4\right)\ga ^\mu
\left[  1-\ga _5\right ] v\left(k_1,\la _3\right) \crn
&\times&   \bar{v}\left(p_1\right)\ga _\mu \left[  g_V^{Z}(e)-{g_A^{Z} (e)}
\ga _5\right ]u\left(p_2\right) \label{Zamplitude}\\
i\mathcal{M}^{W}_{fi}&=& -\left( \fr{g^2}{8M_{W}^2}\right)\bar{u}\left(k_2,\la _4\right)\ga ^\mu
\left[ 1-\ga _5\right ] u\left(k_1,\la _3\right)\crn
&\times&   \bar{v}\left(p_1\right)\ga _\mu \left[  1-\ga _5\right ]v\left(p_2\right)\label{Wamplitude} \\
i\mathcal{M}^{Z'}_{fi}&=& -\left( \fr{g^2}{4\cos^2{\theta_W}M_{Z'}^2}\right)\bar{u}\left(k_2,\la _4\right)
\ga ^\mu \left[  g_V^{Z'}(\nu)-{g_A^{Z'} (\nu)}\ga _5\right ] v\left(k_1,\la _3\right)\crn
&\times&   \bar{v}\left(p_1\right)\ga _\mu \left[  g_V^{Z'}(e)-{g_A^{Z'} (e)}
\ga _5\right ]u\left(p_2\right)\,.\label{Zpamplitude}
\eea

Using Fierz transformation \cite{generalized-Fierz} and summing
over (\ref{Zamplitude}) and (\ref{Wamplitude}) we have
\bea
i\mathcal{M}^{SM}_{fi}&=& -\left( \fr{g^2}{8\cos^2{\theta_W}M_{Z}^2}\right)\bar{u}
\left(k_2,\la _4\right)\ga ^\mu \left[  1-\ga _5\right ] v\left(k_1,\la _3\right) \crn
&\times&   \bar{v}\left(p_1\right)\ga _\mu \left[ C_V(e)-C_A(e)\ga _5\right ]u
\left(p_2\right)  \label{SMamplitude}
\eea
where  $C_V(e)=1+g_V^Z(e)=\fr{1}{2}+2\sin^2{\theta_W}$ and  $C_A(e)=1+g_A^Z(e)=\fr{1}{2}$.

The final amplitude is
\be
\mathcal{M}_{fi}^{331\beta}=\mathcal{M}_{fi}^{SM} + \mathcal{M}_{fi}^{\ga }
+\mathcal{M}_{fi}^{Z'}\,.
\label{331betaamplitude}
\ee

Squaring  (\ref{331betaamplitude}), summing and and taking average over spin of particles in the final and initial states, respectively, we get:
\bea
&&\fr{1}{4}\sum_{s}\left|\mathcal{M}_{fi}\right|^2=\fr{\left(4\pi \al \right)^2 }{\sin^4 {2\theta_W}}
\Bigg\lbrace \fr{\sin^4 {2\theta_W} }{\left(4\pi \al \right)}(\mu_\nu^2 + d_\nu^2 )
\Big[\fr{(p_1\cdot p_2 +m_e^{2})(p_1\cdot k_2+m_e^{2} ) }{s  } \Big]\crn
&&+\Bigg[\bigg(\fr{1}{M_Z^2}\big[ C_V(e)-C_A(e) \big]+\fr{1}{M_{Z^\prime}^2}
\big[g_V^{Z' }(e)-g_A^{Z' }(e) \big]\big[g_V^{Z' }(\nu)+g_A^{Z' }(\nu) \big]\bigg)^2\crn
&&+\bigg( \fr{1}{M_{Z^\prime}^2}\big[g_V^{Z' }(e)+g_A^{Z' }(e) \big]
\big[g_V^{Z' }(\nu)-g_A^{Z' }(\nu) \big]\bigg)^2\Bigg]\times \left(p_1\cdot k_1 \right)\left(p_2\cdot k_2 \right) \crn
&&+\Bigg[ \bigg(\fr{1}{M_Z^2}\big[ C_V(e)+ C_A(e) \big] +\fr{1}{M_{Z^\prime}^2}
 \big[g_V^{Z' }(e)+g_A^{Z' }(e) \big]\big[g_V^{Z' }(\nu)+g_A^{Z' }(\nu) \big]\bigg)^2\crn
&&+\bigg( \fr{1}{M_{Z^\prime}^2} \big[g_V^{Z' }(e)-g_A^{Z' }(e) \big]
\big[g_V^{Z' }(\nu)-g_A^{Z' }(\nu) \big]\bigg)^2\Bigg]
\times \left(p_1\cdot k_2 \right)\left(p_2\cdot k_1 \right) \crn &&+2\Bigg[\fr{1}{2 M_Z^4}
\Big[  \left(C_V(e)\right)^2 -\left(C_A(e)\right)^2 \Big]  +\fr{1}{M_{Z'}^4}
\Big[ \left(g_V^{Z'}(e)\right)^2 -\left(g_A^{Z'}(e)\right)^2 \Big]\crn
&&\times \Big[ \left(g_V^{Z'}(\nu)\right)^2 +\left(g_A^{Z'}(\nu)\right)^2 \Big]  \crn
&&+\fr{2}{M_Z^2 M_{Z'}^2}\left[C_V(e)g_V^{Z'} (e)-C_A(e)g_A^{Z'} ( e)\right]
\left[ \fr{1}{2}g_V^{Z'}( \nu)+\fr{1}{2}(\nu)g_A^{Z'}(  \nu)\right]\Bigg]  \crn
&&\times \left(m_e^{2}\right)\left(k_1\cdot k_2 \right)\Bigg\rbrace \crn
\label{Amplitude}
\eea

Denoting new coefficients defined as follows:
\bea
\mathcal{C}_1&=&\Bigg[\bigg(\fr{1}{M_Z^2}\big[ C_V(e)-C_A(e) \big]+
\fr{1}{M_{Z^\prime}^2} \big[g_V^{Z' }(e)-g_A^{Z' }(e) \big]\big[g_V^{Z' }(\nu)+g_A^{Z' }(\nu) \big]\bigg)^2\crn
&+&\bigg( \fr{1}{M_{Z^\prime}^2}\big[g_V^{Z' }(e)+g_A^{Z' }(e) \big]
\big[g_V^{Z' }(\nu)-g_A^{Z' }(\nu) \big]\bigg)^2\Bigg]  \\
\mathcal{C}_2&=&\Bigg[ \bigg(\fr{1}{M_Z^2}\big[ C_V(e)+ C_A(e) \big]+\fr{1}{M_{Z^\prime}^2}
 \big[g_V^{Z' }(e)+g_A^{Z' }(e) \big]\big[g_V^{Z' }(\nu)+g_A^{Z' }(\nu) \big]\bigg)^2\crn
&+&\bigg( \fr{1}{M_{Z^\prime}^2} \big[g_V^{Z' }(e)-g_A^{Z' }(e) \big]
\big[g_V^{Z' }(\nu)-g_A^{Z' }(\nu) \big]\bigg)^2\Bigg]\\
\mathcal{C}_3&=&2\Bigg[\fr{1}{2 M_Z^4}\Big[  \left(C_V(e)\right)^2 -\left(C_A(e)\right)^2 \Big]  \crn
&+&\fr{1}{M_{Z'}^4}\Big[ \left(g_V^{Z'}(e)\right)^2 -\left(g_A^{Z'}(e)\right)^2 \Big]
\Big[ \left(g_V^{Z'}(\nu)\right)^2 +\left(g_A^{Z'}(\nu)\right)^2 \Big] \crn
&+&\fr{2}{M_Z^2 M_{Z'}^2}\big[C_V(e)g_V^{Z'} (e)-C_A(e)g_A^{Z'} (e)\big]
\big[ \fr{1}{2}g_V^{Z'}( \nu)+\fr{1}{2}g_A^{Z'}(  \nu)\big]\Bigg]\,,
\eea
then Eq. (\ref{Amplitude}) can be rewritten as
\bea
&&\fr{1}{4}\sum_{s}\left|\mathcal{M}_{fi}\right|^2=\fr{\left(4\pi \al \right)^2 }{ \sin^4 {2\theta_W}}
\Bigg\lbrace \fr{\sin^4 {2\theta_W} }{s \left(4\pi \al \right)}\left(\mu_\nu^2 + d_\nu^2 \right)
\left[%\fr{
\left(p_1\cdot p_2 +m_e^{2}\right)\left(p_1\cdot k_2+m_e^{2} \right)
%%{\left(p_1+p_2 \right)^2 }
 \right]
\crn
&&+ \mathcal{C}_1\times \left(p_1\cdot k_1 \right)\left(p_2\cdot k_2 \right)
+\mathcal{C}_2 \times \left(p_1\cdot k_2 \right)\left(p_2\cdot k_1 \right)
+\mathcal{C}_3\left(m_e^{2}\right)\left(k_1\cdot k_2 \right)\Bigg\rbrace\,.
\label{Amplitudef}
\eea
where $ s=\left(p_1+p_2 \right)^2 = \left(k_1+2_2 \right)^2 $.
For future presentation, let us concrete the notations:  four momentum is  $p_i=(E_i,\vec{p}_i), i=1,2$  and $k_i=(E_i,\vec{k}_i), i=1,2$.

\subsection{Stellar energy loss rates}

The  ionized gas in thermal equilibrium %at
with a temperature $T$ and density $\rho$ is suggested to be exist
in the star content.

 Due to Fermi-Dirac distributions, the number densities of the electrons and positrons are given by
\be
n_{\pm}=\int dn_{\pm}=\fr{2}{(2\pi)^3}\int
\fr{d^3p}{\left(e^{\fr{E\mp \mu}{k T}}+1\right)}\,,
\label{numberdensity}
\ee
where $\mu$ is the electron chemical potential.% of the electron.
%The number density of proton is \cbb{determined as} $n_0=n_- - n_+$.
Then, the plasma  mass density %of the plasma
 is  given as:
\be
n_0=n_- - n_+=N \fr{\rho}{\mu_e}
\label{plasmarho}
\ee
where $N$ is Avogadro's number.

The expression for the stellar %energy loss rates
ELRs for
 pair-annihilation process in (\ref{process})
is determined by \cite{Dipole-Eloss-Kerimov1992,neutrino-Eloss-esposito2002,Dipole-Eloss-Heger2009,QSM-Dicus1972}:

\bea
\mathcal{Q}_{\nu\bar{\nu}}^{331\beta}&&=\fr{4}{\left(2\pi\right)^6}\int_{m_e}^{\infty}{\fr{d^3\mathbf{p}_1}{
\left[ e^{\left(E_1-\mu\right)/T}+1\right] }\fr{d^3\mathbf{p}_2}{\left[ e^{\left(E_2+\mu\right)/T}
+1\right] }\left(E_1+E_2\right)}v
\si _{tot}^{331\beta},\crn
&&=\fr{16}{(2\pi)^4}\int_{m_e}^{\infty}{\fr{|\mathbf{p}_1|E_1 dE_1 }{
\left[ e^{\left(E_1-\mu\right)/T}+1\right] }\fr{|\mathbf{p}_2|E_2 dE_2}{\left[ e^{
\left(E_2+\mu\right)/T}+1\right] }\left(E_1+E_2\right)}v
\si _{tot}^{331\beta},
\label{Qloss}
\end{eqnarray}
where $\si _{tot}$ is the %total spin-averaged
process cross-section,   $\left[ \exp\left( {\left(E_{1,2}\pm\mu\right)/T}\right)+1\right]^{-1} $
is the Fermi-Dirac distribution
functions for electron/positron,  %$ e^{\pm} $,
$\mu$ is  the electron  chemical potential,
%for the electron,
$T$ is  the stellar temperature and $v$ is the electron-positron relative
 velocity $\fr{1}{2}[s(s-4m_e^{2})]^{1/2}$ .

The quantity $ E_1E_2v \si _{tot}$ is given by \cite{Dipole-Eloss-Heger2009}
\bea
E_1E_2v
\si _{tot}^{331\beta} &= &\fr{1}{4}\int{\fr{d^3k_1d^3k_2}{\left( 2\pi\right)^{3}2E_3
 \left( 2\pi\right)^{3}2E_4 }} %\left|\mathcal{M}_{\nu\bar\nu}\right|^2
 \left|\mathcal{M}_{fi}\right|^2
 \left( 2\pi\right)^{4}
 \de ^4\left(p_1+p_2-k_1-k_2 \right)\crn
%{E_1E_2v}\si _{Tot}^{331\beta}
&=&\fr{\pi \al ^2 }{3\sin^4 {2\theta_W}}\Bigg\lbrace
\fr{\sin^4\theta_{W}}{2\pi\al }\left(\mu_\nu^2 + d_\nu^2 \right)\left(2m_e^2+p_1\cdot p_2\right)+
\Big(\mathcal{C}_{1}^{331\beta}\Big)\Big[m_e^4\crn
&+&3m_e^2\left(p_1\cdot p_2 \right)+2\left(p_1\cdot p_2 \right)^2  \Big]+12\Big(\mathcal{C}_{2}^{331\beta}\Big)
\Big[m_e^4+m_e^2\left(p_1\cdot p_2 \right)\Big]\Bigg\rbrace \, , \crn
\eea
where
 the coefficients $\mathcal{C}_{1,2}^{331\beta}$  are
 defined as:
\bea
&&\mathcal{C}_{1}^{331\beta}=\dfrac{1}{2 M_Z^4}\Big[ C^2_V(e) +C^2_A(e) \Big]
+\dfrac{1}{M_{Z^\prime}^4}\Big[ \left(g_V^{Z' }(e)\right)^2 +\left(g_A^{Z'}(e)\right)^2 \Big]\crn
&&\times \Big[\left(g_V^{Z'}(\nu)\right)^2 +\left(g_A^{Z'}(\nu)\right)^2 \Big]
+\dfrac{1}{M_Z^2M_{Z^\prime}^2} \big[C_V( e)g_V^{Z' }(e)+C_A(e)g_A^{Z' }(e)\big]\crn
&& \times \big[ g_V^{Z'}(\nu)+g_A^{Z'}(\nu)\big]\,,\crn
%\eea
%\bea
&&\mathcal{C}_{2}^{331\beta}=\dfrac{1}{ 2 M_Z^4}\Big[ C^2_V(e) - C^2_A(e) \Big]
+\dfrac{1}{M_{Z^\prime}^4}\Big[ \left(g_V^{Z' }(e)\right)^2 - \left(g_A^{Z'}(e)\right)^2 \Big]\crn
&&\times \Big[\left(g_V^{Z'}(\nu)\right)^2 +\left(g_A^{Z'}(\nu)\right)^2 \Big] +\dfrac{1}{M_Z^2M_{Z^\prime}^2}
 \big[C_V( e)g_V^{Z' }(e)-C_A(e)g_A^{Z' }(e)\big] \crn
&& \times \big[ g_V^{Z'}(\nu)+g_A^{Z'}(\nu)\big]\,.
\eea

The calculation of the stellar ELRs given by (\ref{Qloss}) can be
fulfill by expressing
in terms of the Fermi integral  defined as
\be
G_n^\pm(\la ,\eta,x)=\la ^{{3+2n}}\int_{\la ^{-1}}^{\infty}{x^{2n+1}
\fr{\sqrt{x^2-\la ^{-2}}}{\left(e^{\left(x\pm\eta\right)}+1\right)}}dx,
\ee
where dimensionless variables $\la$ and $\eta$ are defined as:
\be
\la =\fr{k_{B} T}{m_e},  \hs \eta=\fr{\mu_e}{k_BT}.
\ee
%Taking
Assuming $ k_{B}=1 $ for the Boltzmann constant and from
(\ref{numberdensity}) and  (\ref{plasmarho})  ones have:
\be
G_s^\pm(\la ,\eta,E)=\fr{1}{m_e^{3+2s}}\int_{m_e/T}^{\infty}{E^{2s+1}
\fr{\sqrt{E^2-m_e^{2}}}{\left(e^{\left(E\pm\mu_e\right)/T}+1\right)}}dE\,,
\ee
and
\be
N\fr{\rho}{\mu_e}=\fr{m_e^3}{\pi^2}[G_0^- - G_0^+]\,.
\ee

Hence, the stellar
ELRs  can be expressed as:
\bea
&&\mathcal{Q}^{331\beta}=\fr{\al ^2 m_e^9}{9\pi^3\sin^4 {2\theta_W}}\Bigg\lbrace\fr{3\sin^4{2\theta_{W}}}{2\pi\al
 m_e^2}\left(\mu_\nu^2 + d_\nu^2 \right)\crn
&& \times\bigg[2\Big(G_{-1/2}^-G_{0}^+ + G_{0}^-G_{-1/2}^+\Big)+G_{0}^-G_{1/2}^+ +G_{1/2}^-G_{0}^+\bigg]\crn
&&+4\Big(\mathcal{C}_{1}^{331\beta}\Big) \bigg[5\Big(G_{-1/2}^-G_{0}^+ + G_{0}^-G_{-1/2}^+\Big)
+7\Big(G_{0}^-G_{1/2}^+ +G_{1/2}^-G_{0}^+\Big)\crn
&& -2\Big(G_{1}^-G_{-1/2}^+ +G_{-1/2}^-G_{1}^+\Big)+8\Big(G_{1}^-G_{1/2}^+ +G_{1/2}^-G_{1}^+\Big) \bigg]\crn
&&+36\Big(\mathcal{C}_{2}^{331\beta}\Big)\bigg[G_{-1/2}^-G_{0}^+ + G_{0}^-G_{-1/2}^+ +G_{0}^-G_{1/2}^+
 +G_{1/2}^-G_{0}^+\bigg]\Bigg\rbrace.
\label{emissivity}
\eea

To investigate the effects of dipole moments and new contribution of $Z'$ boson on the
ELR we have
 to evaluate the relative correction for the star %energy loss rate
 ELR  between 3-3-1$\beta$ model and that
  of the SM  ($\mathcal{Q}^{SM}$) \cite{QSM-Dicus1972} given as:
\bea
&&\mathcal{Q}^{SM}=\fr{G_F^2m_e^9}{18 \pi}  \Biggl\{ (7C_V(e)^2-2C_A(e)^2)[G_0^- G_{1/2}^+
 + G_{-1/2}^- G_{0}^+    ]
\crn
&&+ 9C_V(e)^2[G_{1/2}^- G_{0}^+ + G_{0}^- G_{1/2}^+    ] + (C_V(e)^2+C_A(e)^2)[4G_{1}^{-} G_{1/2}^{+}
+G_{1/2}^{-} G_{1}^{+} \crn
&&- G_{1}^{-} G_{-1/2}^{+} - G_{1/2}^{-} G_{0}^{+} - G_{0}^{-} G_{1/2}^{+} -
G_{-1/2}^{-} G_{1}^{+} ]  \Biggr\}\,, %\crn
\eea
Hence
\be
\fr{\de  \mathcal{Q}^{331\beta} }{\mathcal{Q}^{SM}}=\fr{\mathcal{Q}^{331\beta}
\left(\mu_{\nu}, d_{\nu},\beta,M_{Z^{\prime}},\eta  \right) - \mathcal{Q}^{SM}
\left(\eta \right)  }{\mathcal{Q}^{SM}\left(\eta \right) }.
\label{QCorrection}
\ee

Since the functions $G^{\pm}(\la , \eta)$ can only be defined analytically in some limiting
regions of parameters $\la  $ and $\eta$ therefore we will investigate the correction (\ref{QCorrection}) in
five regions

\subsubsection{Region I:  $\la \ll 1$ and $ \eta \ll 1/\la$}

In this region, temperature and densities
vary between $3\times10^8 \, \textrm{K}  \le \textrm{T} \le 3\times10^9\, \textrm{K}$ and $\rho \le 10^5 gr/cm^3$, respectively.

The Fermi integral is

\be
G_n^\pm(\la ,\eta,x)=\la ^{{3+2n}}\int_{\la ^{-1}}^{\infty}{x^{2n+1}\fr{\sqrt{x^2
-\la ^{-2}}}{\left(e^{\left(x\pm\eta\right)}+1\right)}} dx \,.
\ee

Changing  variable  $ x=z+ \la ^{-1} $ yields
\be
G_n^{\pm}=\sqrt{2}\la ^{3/2}e^{-\la ^{-1}}e^{\mp \eta}\int_0^\infty
dz (\la  z+ 1)^{2n+1}z^{1/2}\left(1+\fr{\la  z}{2}\right)^{1/2}e^{-z}
\ee

For every $\la  $ satisfying $\la  z  \approx \ep$ we have
\[
(\la  z + 1)^{2n+1} \hs [1+ (2n+1)\la  z], \hs   \left(1+\fr{\la  z}{2}\right)^{1/2} \approx 1+\fr{\la  z}{4}
\]
Neglecting the
second order in $\la z$ ones get
\bea
G_n^{\pm}&&=\sqrt{2}\la ^{3/2}e^{-\la ^{-1}}e^{\mp \eta}\left[ \int_0^\infty dz
 z^{1/2}e^{-z}+\left(2n+\fr{5}{4}\right)\la  \int_0^\infty dz  z^{3/2}e^{-z} \right]  \crn
&&=\sqrt{2}\la ^{3/2}e^{-\la ^{-1}}e^{\mp \eta}\left[  \Ga (3/2)+\left(2n+\fr{5}{4}\la \right)\Ga (5/2) \right] \crn
&&=\sqrt{\fr{\pi}{2}}\la ^{3/2}e^{-\la ^{-1}}e^{\mp \eta}\left[1+\fr{3}{2}\left(2n+\fr{5}{4}\right)\la  \right]\,,
\end{eqnarray}
where $\Ga (n)=\int_0^\infty z^{n-1} e^{-z} dz$.

In the  case
$1 \ll \la ^{-1}$ ones have
\be
G_0^{\pm}\approx\sqrt{\fr{\pi}{2}}\la ^{3/2}e^{-\la ^{-1}}e^{\mp \eta}
\ee
and
\be
G_n^-\approx G_0^-= \left(\fr{\pi}{2}\right)^{\fr{1}{2}}\la ^{\fr{3}{2}}e^{-1/\la }e^{\eta}.
\ee

Hence, we get
\bea
\mathcal{Q}_I^{331\beta}&&=\fr{\al ^2 m_e^9}{\pi^3\sin^4 {2\theta_W}}
\Bigg[\fr{\sin^4{2\theta_{W}}}{\pi\al
 m_e^2}\left(\mu_\nu^2 + d_\nu^2 \right)   +  16(\mathcal{C}_1^{331\beta}+\mathcal{C}_2^{331\beta}) \Bigg]  G_0^-G_0^+ \crn
&&=\fr{\al ^2 m_e^9}{\pi^3\sin^4 {2\theta_W}}\Bigg[\fr{\sin^4{2\theta_{W}}}{\pi\al  m_e^2}\left(\mu_\nu^2 + d_\nu^2 \right)  +  16(\mathcal{C}_1^{331\beta}+\mathcal{C}_2^{331\beta}) \Bigg] \crn
&&\times \left( \fr{\pi}{2}\right)^{3/2} \left(\fr{T}{m_e}\right)^3 e^{-2m_e/T}
\label{Qloss-I}
\eea
and
\bea
\mathcal{Q}_I^{SM}&&=\fr{G_F^2 m_e^9}{18 \pi^5}\times 36 C^2_V(e) G_0^-G_0^+ \crn
&&=\fr{\al ^2 m_e^9}{\pi^3 \sin^4{2\theta_W}}\fr{16}{ m_Z^4} C_V^2(e)G_0^- G_0^+
\label{QlossSM-I}
\eea

Then, the  correction  is given by:

\be
\fr{\de  \mathcal{Q}_I^{331\beta}}{\mathcal{Q}_I^{SM}}=
\fr{\Bigg[\fr{\sin^4{2\theta_{W}}}{\pi\al  m_e^2}\left(\mu_\nu^2 +
 d_\nu^2 \right)  +  16(\mathcal{C}_1^{331\beta}+\mathcal{C}_2^{331\beta})
  \Bigg]- \fr{16}{ m_Z^4} C_V^2(e)}{\fr{16}{ m_Z^4} C_V^2(e)}
\label{QCorrection-I}
\ee

\subsubsection{Region II: $\la  \ll 1$ and $ \fr{1}{\la } \ll \eta \ll \fr{2}{\la } $}

This region is  nonrelativistic and mildly degenerate, with temperatures $T\le10^8$ K  and
densities between $10^4 gr/cm^3\,  \le \rho \le \,  10^6 gr/cm^3$.

The  Fermi integrals satisfy
the following conditions \cite{QBSM-Hernandez,QBSM-Hernandez,QSM-Dicus1972}:  $G_0^-\gg G_0^+$ and $G_n^- \approx G_0^-$ and

\be
G_n^+\approx G_0^+=\left(\fr{\pi}{2}\right)^{\fr{1}{2}}\la ^{\fr{3}{2}}
e^{-1/\la }e^{-\eta}, \qquad G_n^-\approx G_0^-=\left(\fr{\rho}{\mu_e}\right)\fr{\pi^2}{m_e^3}N_A.
\ee

Then ones have:	
\bea
\mathcal{Q}_{II}^{331\beta}&&=\fr{\al ^2 m_e^9}{\pi^3\sin^4 {2\theta_W}}\Bigg[\fr{\sin^4{2\theta_{W}}}{\pi\al
 m_e^2}\left(\mu_\nu^2 + d_\nu^2 \right)  +  16(\mathcal{C}_1^{331\beta}
 +\mathcal{C}_2^{331\beta})\Bigg] G_0^-G_0^+ \crn
&&=\fr{\al ^2 m_e^6}{\pi\sin^4 {2\theta_W}}     \left( \fr{\rho}{\mu_e}N_A\right)
 \left(\fr{\pi}{2}\right)^{1/2}
\left(\fr{T}{m_e}\right)^{3/2}  e^{-(m_e+\mu_e)/T} \crn
&& \times \Bigg[\fr{\sin^4{2\theta_{W}}}{\pi\al  m_e^2}\left(\mu_\nu^2 + d_\nu^2 \right)  +  16(\mathcal{C}_1^{331\beta}+\mathcal{C}_2^{331\beta})\Bigg]
\label{Qloss-II}
\eea
and
\bea
\mathcal{Q}_{II}^{SM}&&=\fr{G_F^2 m_e^9}{18 \pi^5}\times 36 C^2_V(e) G_0^-G_0^+ \crn
&&=\fr{\al ^2 m_e^9}{\pi^3 \sin^4{2\theta_W}}\fr{16}{ m_Z^4} C_V^2(e)G_0^- G_0^+ \crn
&&=\fr{\al ^2 m_e^9}{\pi^3 \sin^4{2\theta_W}}\fr{16}{ m_Z^4} C_V^2(e)\left(
\fr{\rho}{\mu_e}N_A \fr{\pi^2}{m_e^3}\right)  \left(\fr{\pi}{2}\right)^{1/2}
\left(\fr{T}{m_e}\right)^{3/2}  e^{-(m_e+\mu_e)/T}\,.
\label{QlossSM-II}
\eea

Therefore the correction is  given as exactly as in (\ref{QCorrection-I}) which is equal as in region I.

\subsubsection{Region III: $\la  \ll 1$ and\, $ 1 \ll \la \eta $}

The considered region represents the relativistic and degenerate case and is valid for temperatures
$T > 6\times10^7$ K and densities $\rho > 10^7 gr/cm^3$. The Fermi integrals result:
\be
G_n^+\approx G_0^+=\left(\fr{\pi}{2}\right)^{\fr{1}{2}}\la ^{\fr{3}{2}}e^{-1/\la }e^{-\eta},
\hs G_n^-=\left(\fr{3}{2n+3}\right)\left(\la \eta\right)^{2n}G_0^-.
\ee

Then to highest power in $ \la \eta $ ones have
\bea
\mathcal{Q}_{III}^{331\beta}&&=\fr{8\al ^2 m_e^9} {3\pi^3 \sin^4{2\theta_{W}}}
 \mathcal{C}_1^{331\beta}G_1^- G_0^+
\crn
&&=\fr{8\al ^2 m_e^9} {3 \pi^3 \sin^4{2\theta_{W}}}
\fr{3}{5}\left(\fr{1}{m_e} \fr{\mu_e}{T}\right)^2
\left(\fr{\rho}{\mu_e} \right)\fr{\pi^2}{\mu^3_e}N_A
\left(\fr{\pi}{2}\right)^{1/2}\left( \fr{T}{m_e}\right)^{3/2}
\crn
&&\times e^{-({m_{e}+\mu_e})/{T}}  \mathcal{C}_1^{331\beta},
\label{Qloss-III}
\eea
and
\bea
\mathcal{Q}^{SM}=\fr{4}{3}\fr{\al ^2 m_e^9}{\pi^3
\sin^4{2\theta_W}m^4_Z}[ C_V^2(e)+C^2_A(e)]G_1^-G_0^+\,.
\label{QlossSM-III}
\end{eqnarray}
The relativistic correction is given by:
\bea
\fr{\de  \mathcal{Q}^{331\beta}_{III} }{\mathcal{Q}^{SM}_{III}}&=&
\fr{2 m^4_Z\mathcal{C}_1^{331\beta} }{ [C_V^2(e)+C^2_A(e)] }-1\,.
\label{QCorrection-III}
\eea

Since the approximation for this region only considers the terms of dominant powers,
so there is no dependence on the AMM and/or EDM of the neutrino.

\subsubsection{Region IV: $\la  \gg 1$ and $ \eta \ll 1 $}

The relativistic and nondegenerate case holds for densities $\rho>10^7 gr/cm^3$. In this
region we may ignore the chemical potential. Considering the dominance of the highest orders in $\la $, we get
\be
G_n^\pm\approx \la ^{2n+3} \Ga \left(2n+3\right)\sum_{S=1}^{\infty}\fr{\left(-1\right)^{S+1}}{S^{2n+3}}.
\ee

Then, the stellar ELRs for this region is given by
\bea
\mathcal{Q}_{IV}^{331\beta}&&=\fr{64 \al ^2 m_e^9} {9 \pi^3
 \sin^4{2\theta_{W}}} \mathcal{C}_1^{331\beta} G_1^- G_{1/2}^+ \crn
&&=\fr{64 \al ^2 m_e^9} {9 \pi^3 \sin^4{2\theta_{W}}} \mathcal{C}_1^{331\beta}
 \left(\fr{T}{m_e}\right)^9\Ga (5)\xi(5)\Ga (4)\xi(4)
\label{Qloss-IV}
\eea
where $\Ga $ and $\xi$ are the Gamma % function
and Riemann zeta functions, respectively.

For the SM the ELR is given by:
\bea
\mathcal{Q}^{SM}=\fr{32 \al ^2 m_e^9} {9 \pi^3 \sin^4{2\theta_{W}} m^4_Z}(C^2_V(e)
+C^2_A(e)) G_1^- G_{1/2}^+\,.
\label{QlossSM-IV}
\eea

Thus the relativistic correction is the same as in the previous region, namely
\bea
\fr{\de  \mathcal{Q}^{331\beta}_{IV} }{\mathcal{Q}^{SM}_{IV}}&=&
\fr{2 m_Z^4 \mathcal{C}_1^{331\beta}}{[C_V^2(e)+C^2_A(e)]}-1.
\label{QCorrection-IV}
\eea

\subsubsection{Region V: $\la  \gg 1$ and $ \eta \gg 1 $}

This degenerate relativistic region holds for densities greater than $\rho > 10^8 gr/cm^3$
with temperatures of $ T\approx 10^{10}$ K at the lowest density, extendable to a range
between $10^{10}$ K and $10^{11}$ K at a density of $\rho > 10^{10} gr/cm^3$. Here $G_n^-\gg G_n^+$, then
\be
G_n^+\approx \la ^{2n+3} \left(2n+2\right)! e^{-\eta} \hs G_n^-\approx \left(\fr{3}{2n+3}\right)\left(\la \eta\right)^{2n}\left(\fr{\rho}{\mu_e}\right)\fr{\pi^2}{m_e^3}N_A.
\ee

Restricting the calculation to the higher powers in $\la \eta$, the stellar ELRs result:
\bea
\mathcal{Q}^{331\beta}_{V}&&=\fr{32\al ^2 m_e^9}{9\pi^3 \sin^4{2\theta_{W}}}
 \mathcal{C}_1^{331\beta}G_1^- G_{1/2}^+ \crn
&&
=\fr{64\al ^2 m_e^6}{5\pi \sin^4{2\theta_{W}}} \mathcal{C}_1^{331\beta}
\left(\fr{\rho}{\mu_e} N_{A}\right)\left( \fr{T}{m_e}\right)^{6}\left( \fr{\mu_e}{T}\right)^{2}e^{-\mu_e/{T}}
\label{Qloss-V}
\eea
and
\bea
\mathcal{Q}^{SM}_{V}&&=\fr{32\al ^2m_e^9}{9\pi^3 \sin^4{2\theta_{W}} m_Z^4}
 [C_V^2(e)+ C_A^2(e)]G_1^- G_{1/2}^+ \crn
\label{QlossSM-V}
\eea
and  the relativistic correction is given by
\bea
\fr{\de  \mathcal{Q}^{331\beta}_{V} }{\mathcal{Q}^{SM}_{V}}&=&
\fr{2 m_Z^4 \mathcal{C}_1^{331\beta}}{[C_V^2(e)+C^2_A(e)]}-1\, .
\label{QCorrection-V}
\eea
The above result is  equal to that  in Region III. Again, it becomes clear that there is
an indistinguishability of treating with nondegenerate
or degenerate electrons.

\section{Numerical analysis}
\label{result}

Let us  investigate the pair annihilation neutrino ELR  in the context of
the $SU(3)_C\times SU(3)_L\times U(1)_X$ models. The process in (\ref{process})
is one of the main mechanisms of neutrino pair production relevant for the neutrino luminosity.
We investigate for both degenerate and nondegenerate Fermi gas. In the context of beyond Standard Model,
at loop level the non-vanishing of AMM and EDM of neutrinos and the appearance of new neural gauge
 bosons with V-A interaction can contribute to the process of energy loss.

\begin{figure}[ht!]
\centering
\includegraphics[scale=.75]{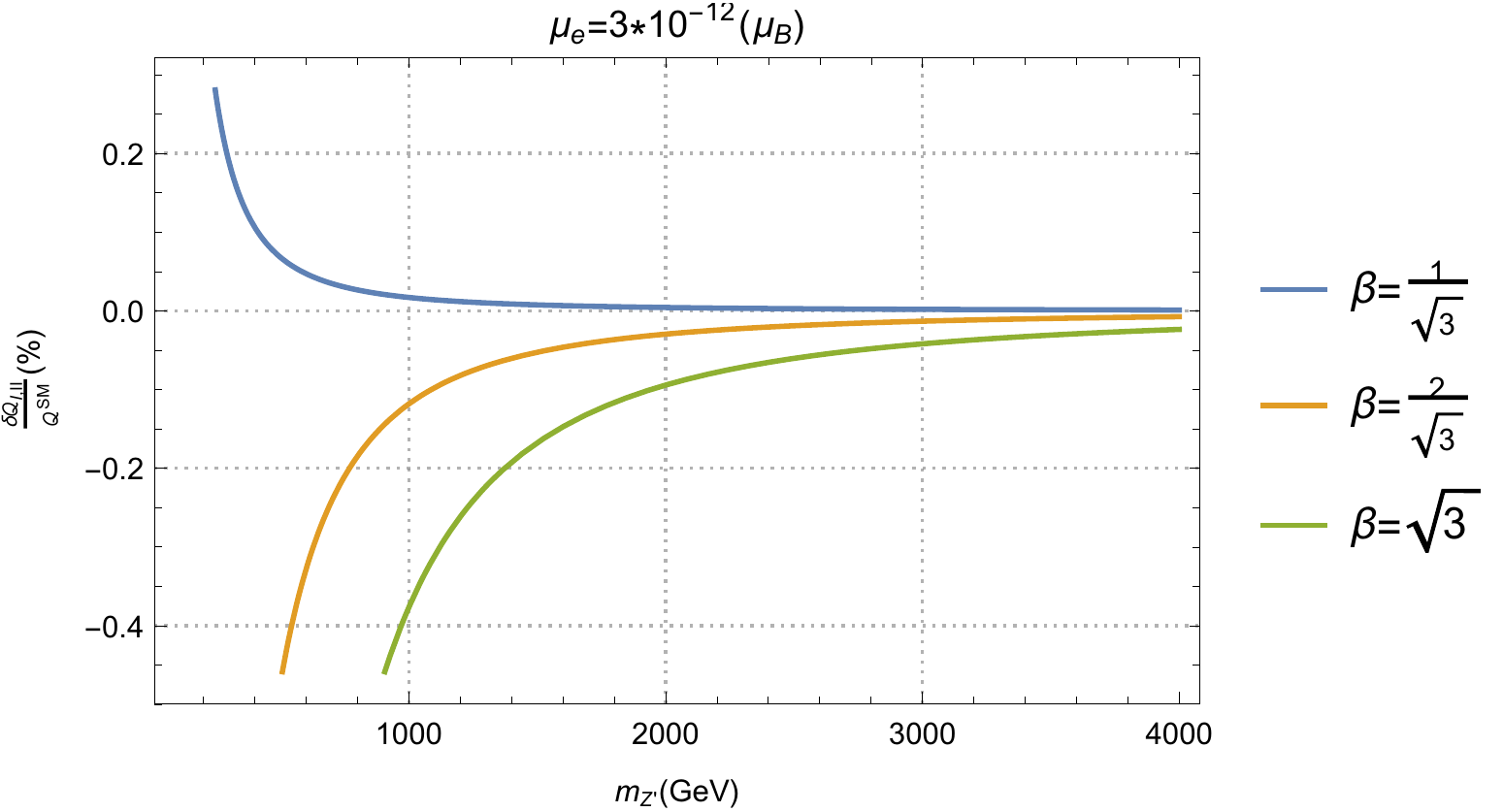}

\caption{Q correction regions I,II with $\beta>0$ }
\label{QCorrection-12a}
\end{figure}

\begin{figure}[ht!]
	\centering
	
	\includegraphics[scale=.75]{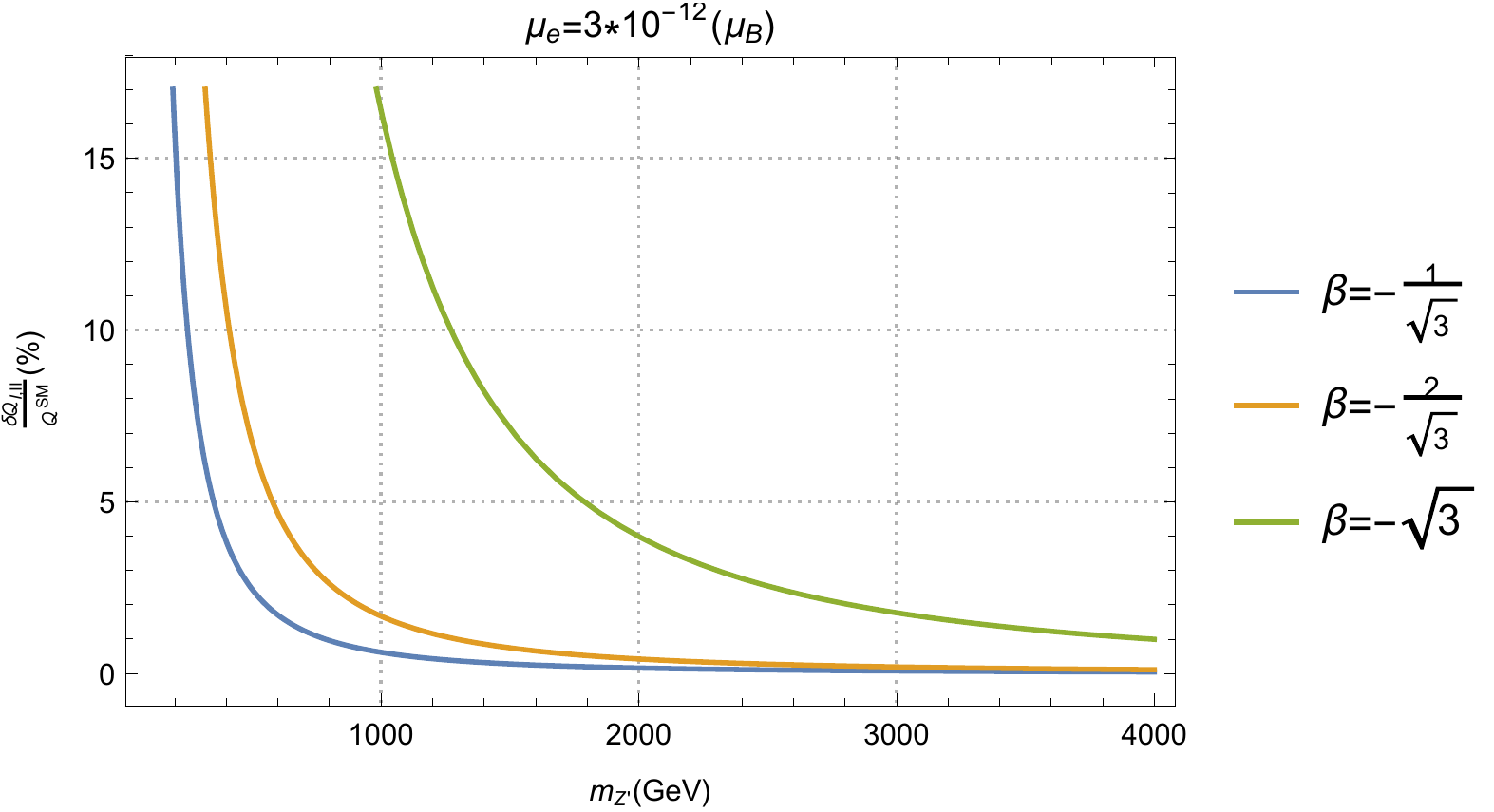}
	\caption{Q correction regions I,II with $\beta<0$}
	\label{QCorrection-12b}
\end{figure}

\begin{figure}[ht!]
\centering
\includegraphics[scale=.75]{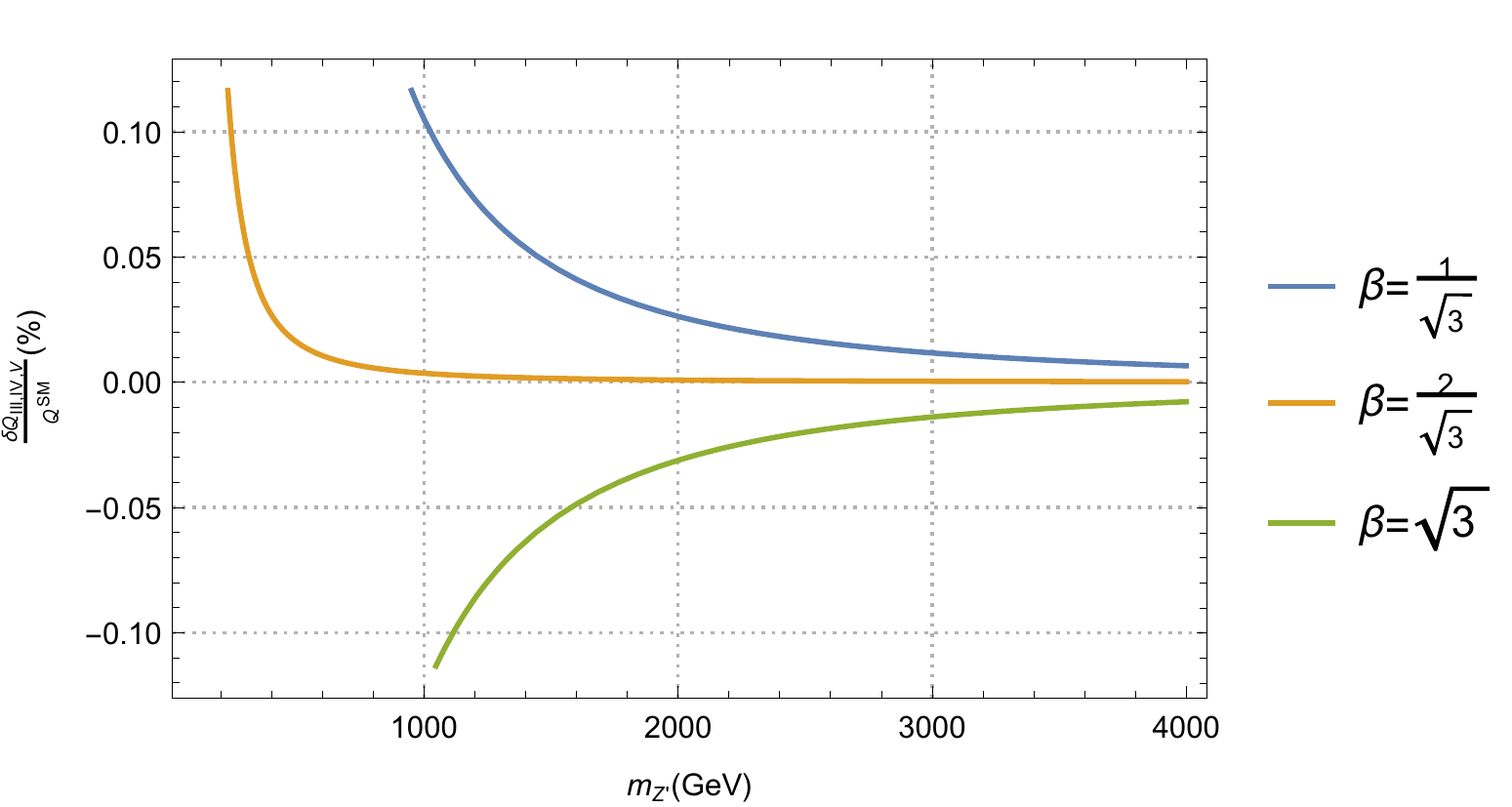}

\caption{Q correction regions III,IV,V  with $\beta>0$}
\label{QCorrection-345a}
\end{figure}

\begin{figure}[ht!]
	\centering
	
	\includegraphics[scale=.75]{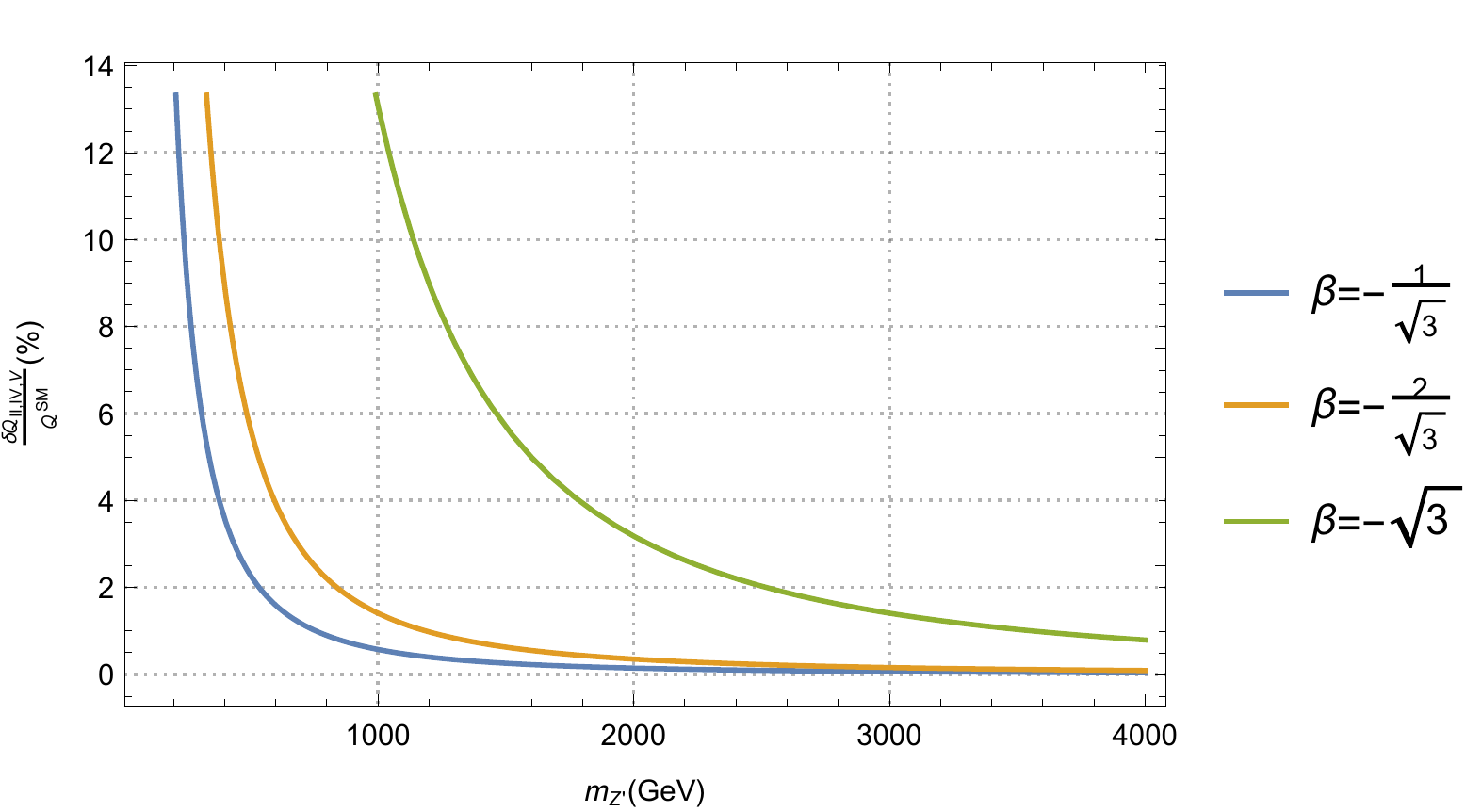}
	\caption{Q correction regions III,IV,V  with $\beta<0$}
	\label{QCorrection-345b}
\end{figure}

Before investigate the effect of temperature and densities on the ELR
we will investigate
the rate of correction of 3-3-1 models compared with the SM because with the same value of temperature
 and densities the parameters that distinguish models is the value of parameter of the model
 (the parameter $\beta$) and the mass of the new gauge bosons $m_{Z'}$.
 We plot the $\mathcal{Q}$
  correction  for regions I, II in Figs.\ref{QCorrection-12a} and \ref{QCorrection-12b}.
and for regions II, IV, V in Figs.\ref{QCorrection-345a} and \ref{QCorrection-345b}

The correction is plotted for all values of parameter $\beta$. For negative value of $\beta$
the correction is up to $15\%$ while positive value of $\beta$ give very small correction ($0.2\%$).
 For all negative $\beta$, the correction decreases with the increase of the mass of Z' boson.
  For $\beta=-\fr{1}{\sqrt{3}},-\fr{2}{\sqrt{3}}$, the correction is approximated to 0\%
  when $m_{Z'}\geq 2000$ GeV and only increase significantly when $m_{Z'}\approx 200$ GeV
   which is around the energy region of the $Z$ boson. In the case $\beta=-\sqrt{3}$, the correction is higher from 1\% up to 15\% in the mass range from 4000 GeV to 1000 GeV.
This case is more interesting when showing that  the  contribution of the $Z'$ boson in the 3-3-1 model
is distinguishable with the SM. Therefore put constraints on the mass range of the $Z'$ boson which is
 in agreement with searching mass range for the $Z'$ boson at LHC \cite{PDG2016,Z' resonance}.
In the followings detail numerical analysis, we will work with the case where $\beta=-\sqrt{3}$.

\begin{figure}[ht!]
\centering
\includegraphics[scale=.75]{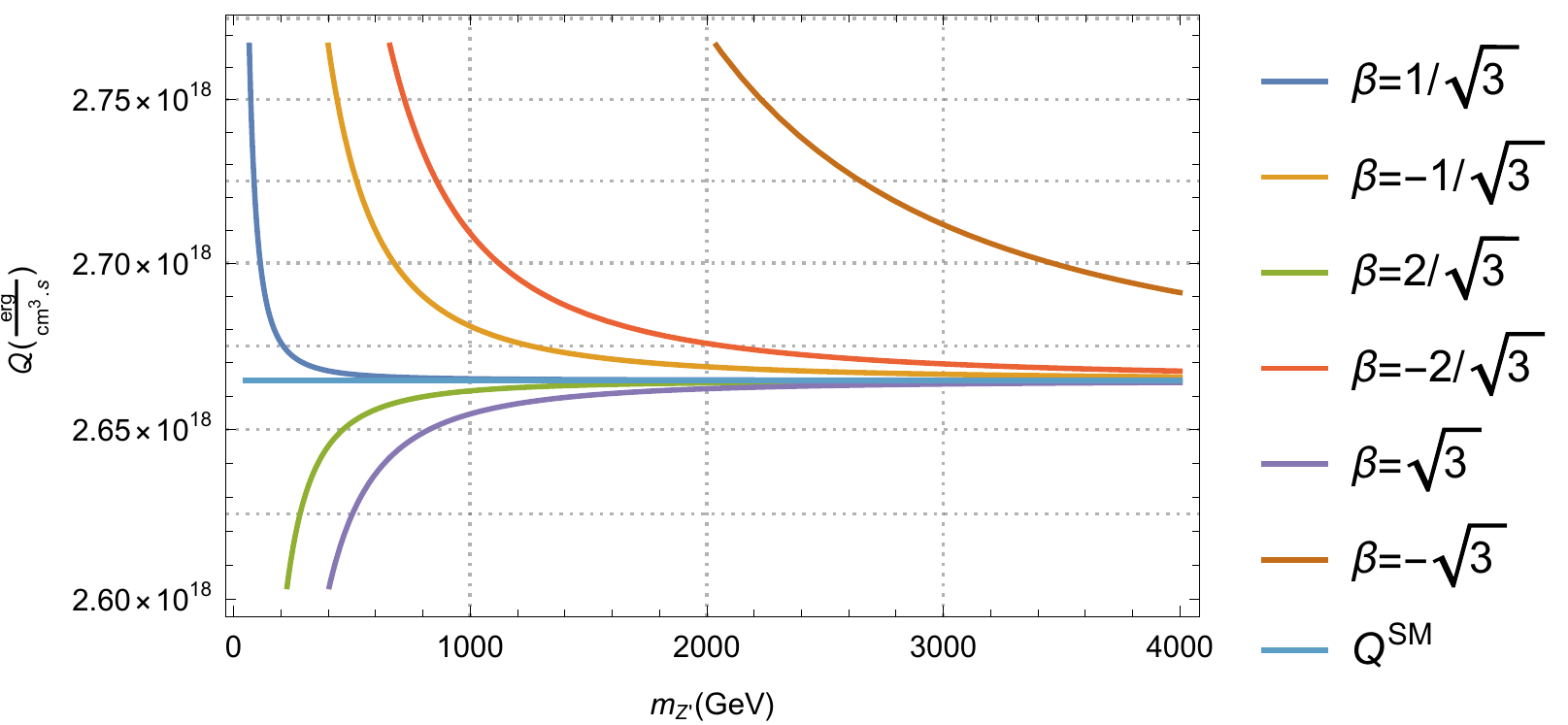}
\caption{Energy loss rate for region I as a function of $m_{Z'}$}
\label{QImZP}
\end{figure}

\begin{figure}[ht!]
\centering
\includegraphics[scale=.75]{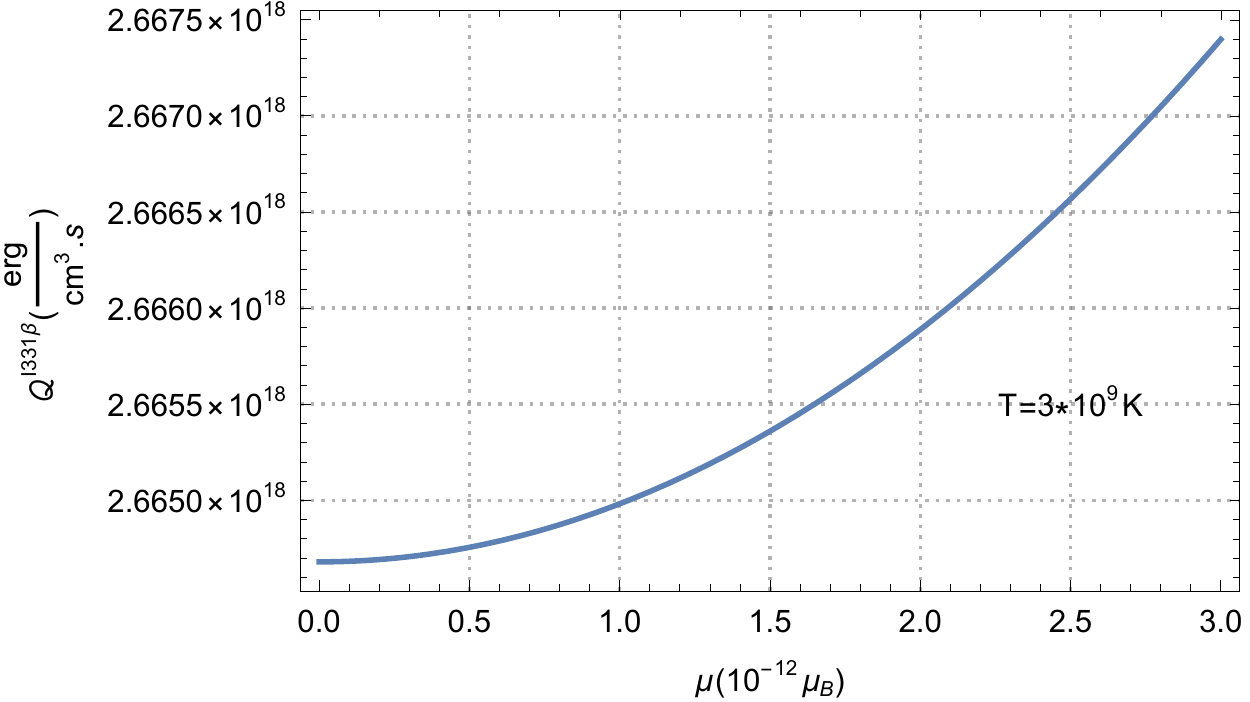}
\caption{Energy loss rate for region I as a function of magnetic dipole moment}
\label{QI331mu}
\end{figure}

In Fig.\ref{QImZP},  we plot
%  investigate
 the energy loss as a function of $M_{Z'}$ and compare it with those in  the SM.
The value of temperature is set $T=3*10^9 $K and $\mu_\nu=3*10^{-12}(\mu_B)$. We plot the contribution of dipole moment and gauge boson $Z'$ separately. In Fig.\ref{QI331mu} we plot
the energy loss rate as a function of magnetic moment.

\begin{figure}[ht!]
\centering
\includegraphics[scale=.75]{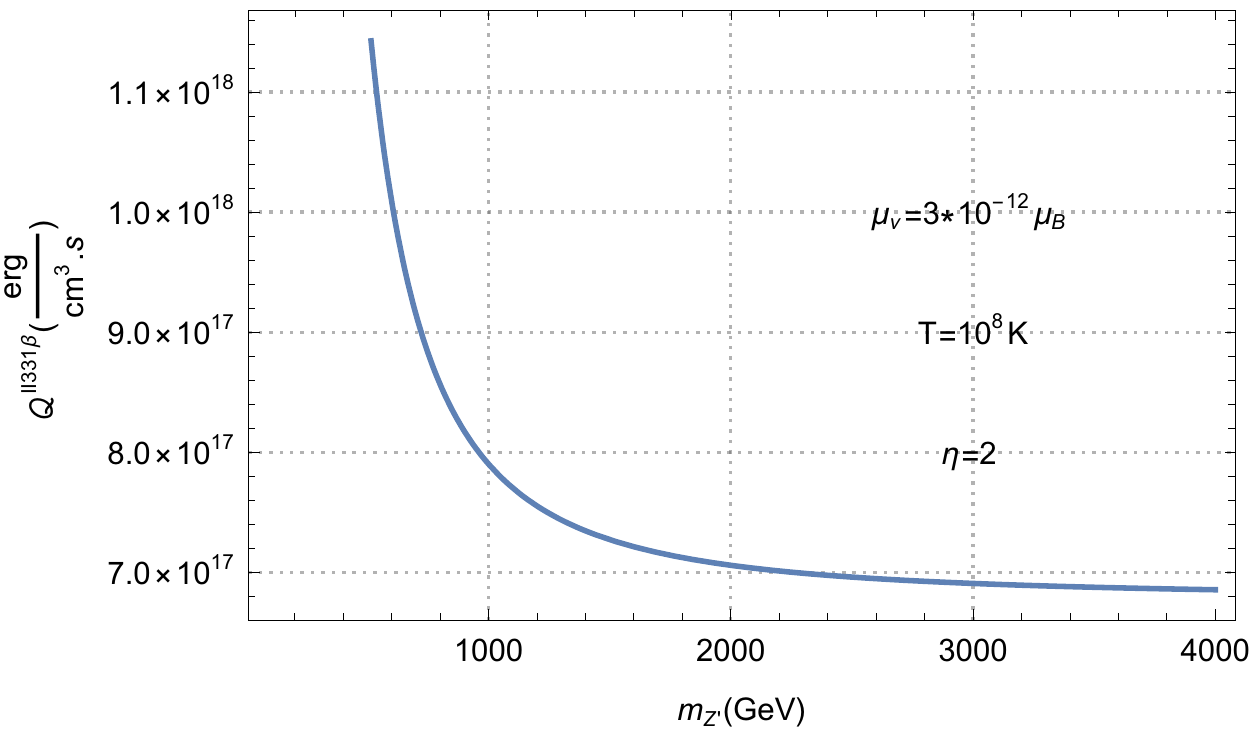}
\caption{Energy loss rate for region II as  a function of $m_{Z'}$ }
\label{QII331mZP}
\end{figure}

\begin{figure}[ht!]
\centering
\includegraphics[scale=.75]{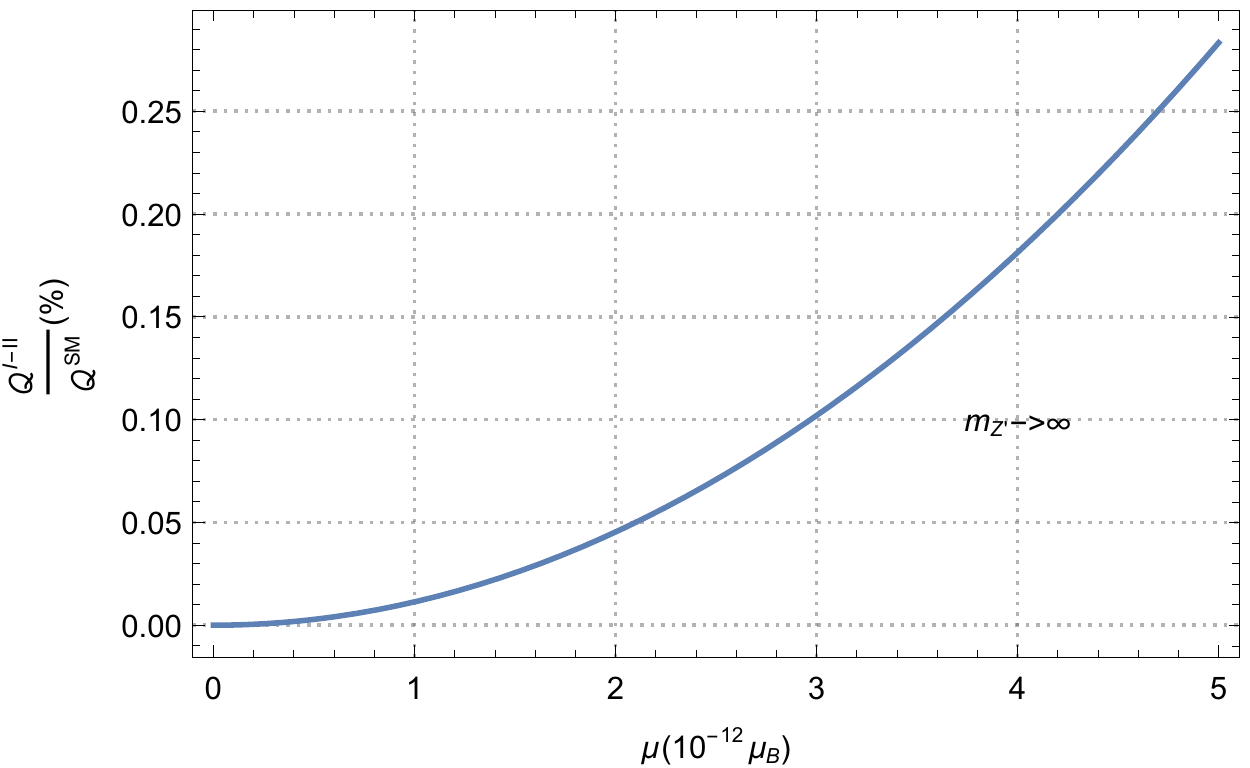}
\caption{Energy loss rate for region I as a function of magnetic dipole moment}
\label{QCor12mbs3mZpinfFig}
\end{figure}

To investigate the effect of only of the dipole moment we take the limit where the value of $m_{Z'}$
very large ($\infty$). In this limit the contribution of the  gauge boson $Z'$ is small. The magnitude of $\mathcal{Q}$
increase with the value of $\mu_\nu$, however with current bound $\mu_\nu <3*10^{-12}\mu_B$ the
 value of $\mathcal{Q}< 2.6675*10^{18}$ which is approximate the magnitude of those in the SM as in Fig.\ref{QImZP}
 or the $\mathcal{Q}$ correction for region I-II is less than 0.1\%( Fig.\ref{QCor12mbs3mZpinfFig})
 comparing with 14\% of the combine contribution of both dipole moment and the $Z'$  boson.
 Hence the effect of dipole moment is not clear in this case.

\begin{figure}[ht!]
\centering
\includegraphics[scale=.75]{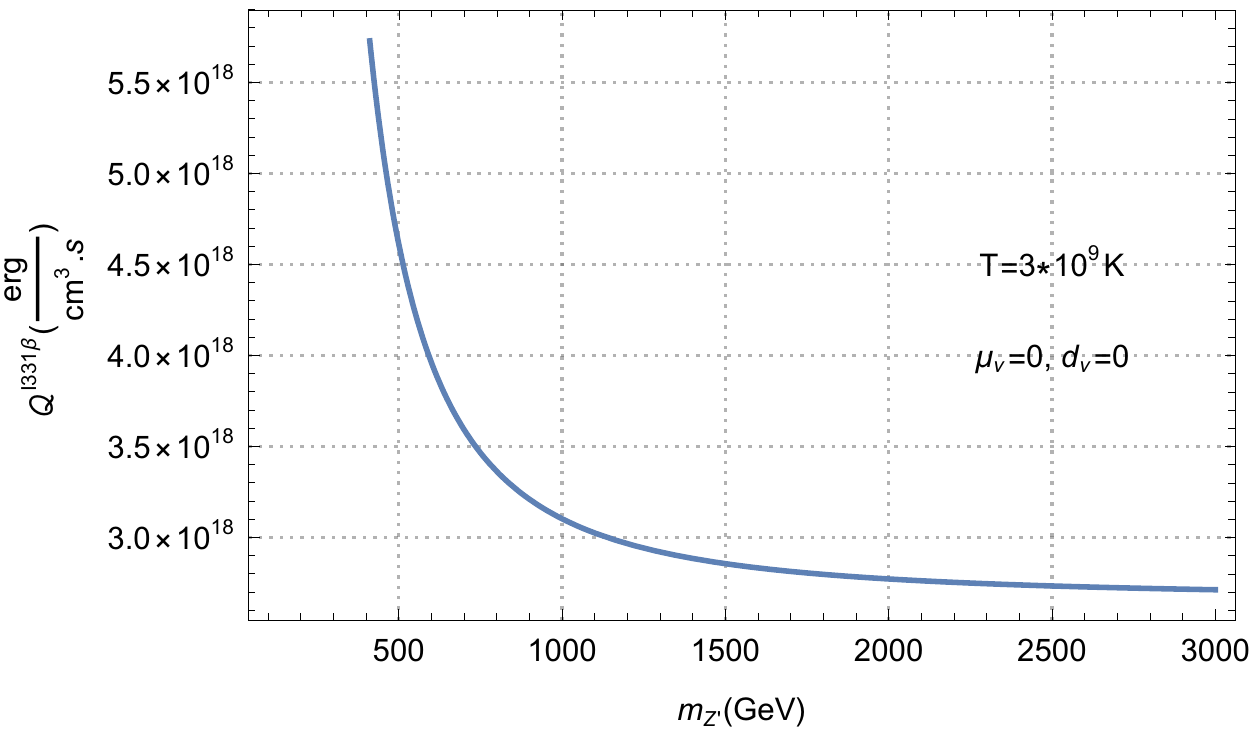}
\caption{Energy loss rate for region I as a function of $m_{Z'}$, for  $\mu_\nu=0, d_\nu=0$,  $T=3*10^9 $ K}
\label{QI331mu0mZPFig}
\end{figure}

 The Fig.\ref{QI331mu0mZPFig} is the case for $\mu_\nu=0, d_\nu=0$ and the value of temperature   $T=3*10^9 $ K.
Plot $\mathcal{Q}$ as function of $m_{Z'}$,  for mass range of $m_{Z'}< 2000$ GeV the magnitude
of energy loss $\mathcal{Q} \approx 2.8 *10^{18}$ therefore the effect of gauge boson $Z'$ is much  stronger than that of dipole moment.

\begin{figure}[ht!]
\centering
\includegraphics[scale=.75]{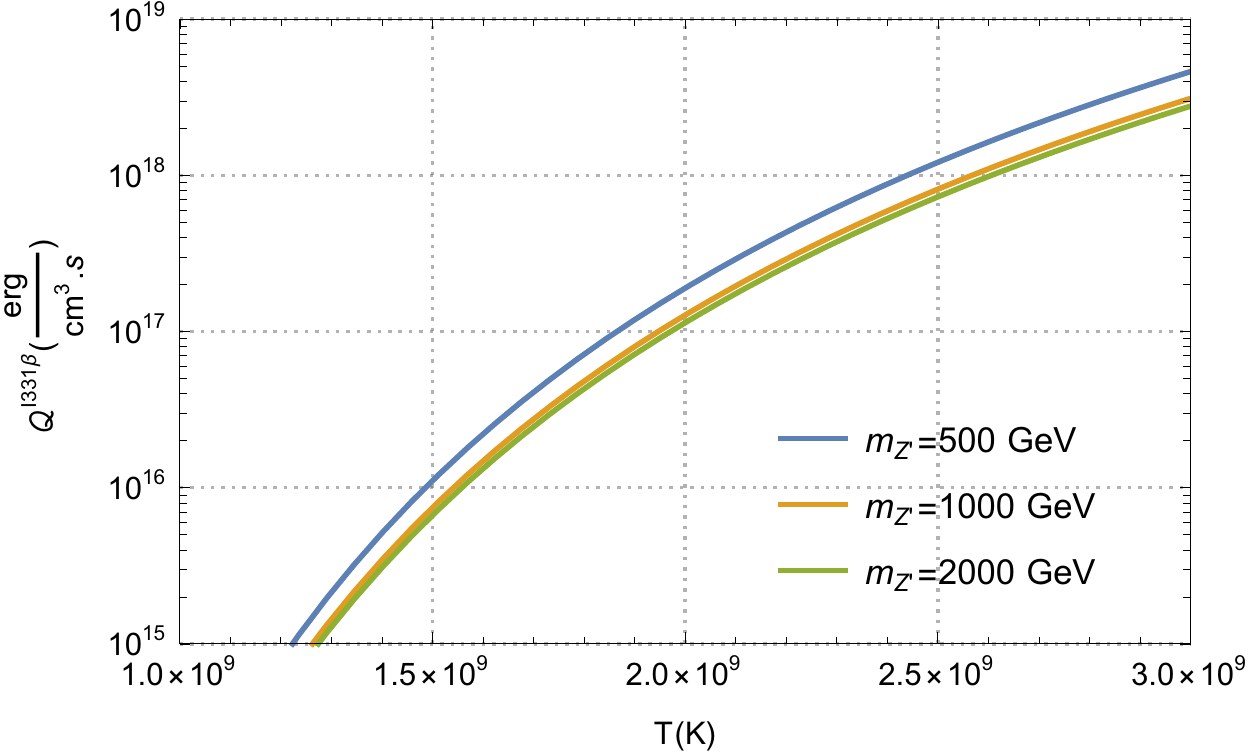}
\caption{Energy loss rate for region I as a function of temperature}
\label{QITK}
\end{figure}

\begin{figure}[ht!]
\centering
\includegraphics[scale=.75]{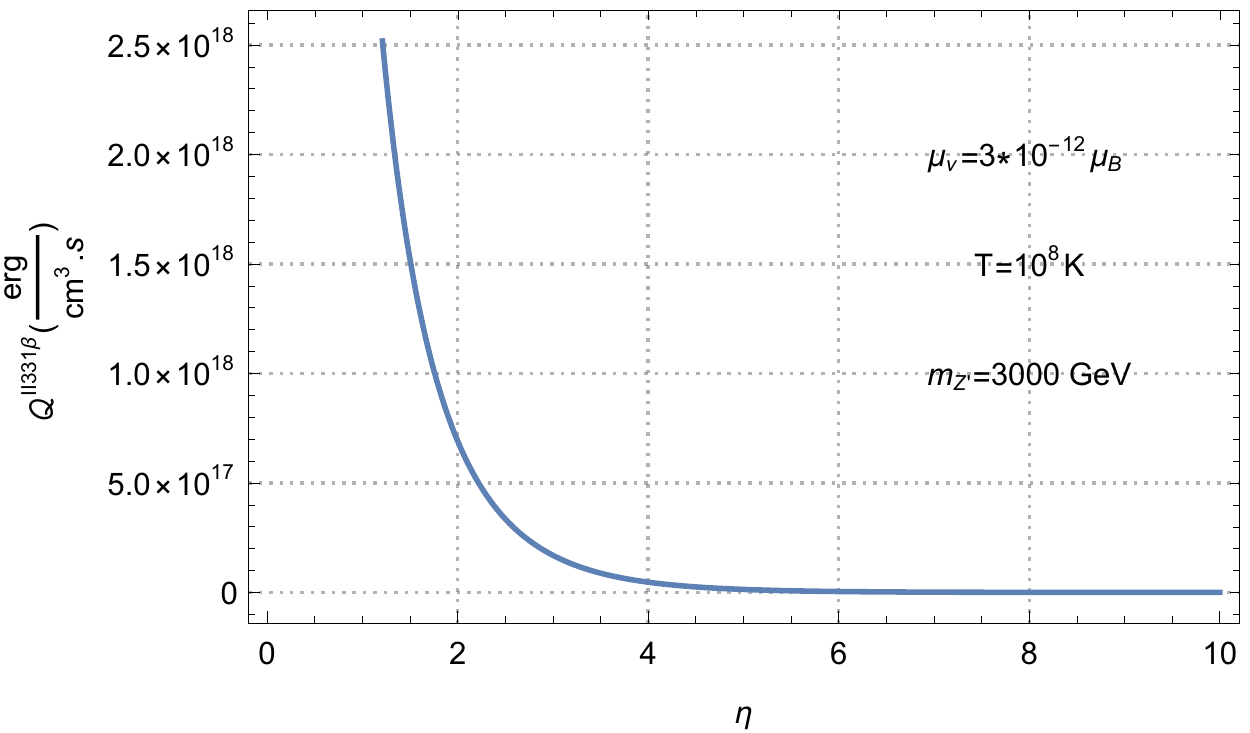}
\caption{Energy loss rate for region II as a function of $\eta$ }
\label{QII331eta}
\end{figure}

\begin{figure}[ht!]
\centering
\includegraphics[scale=.65]{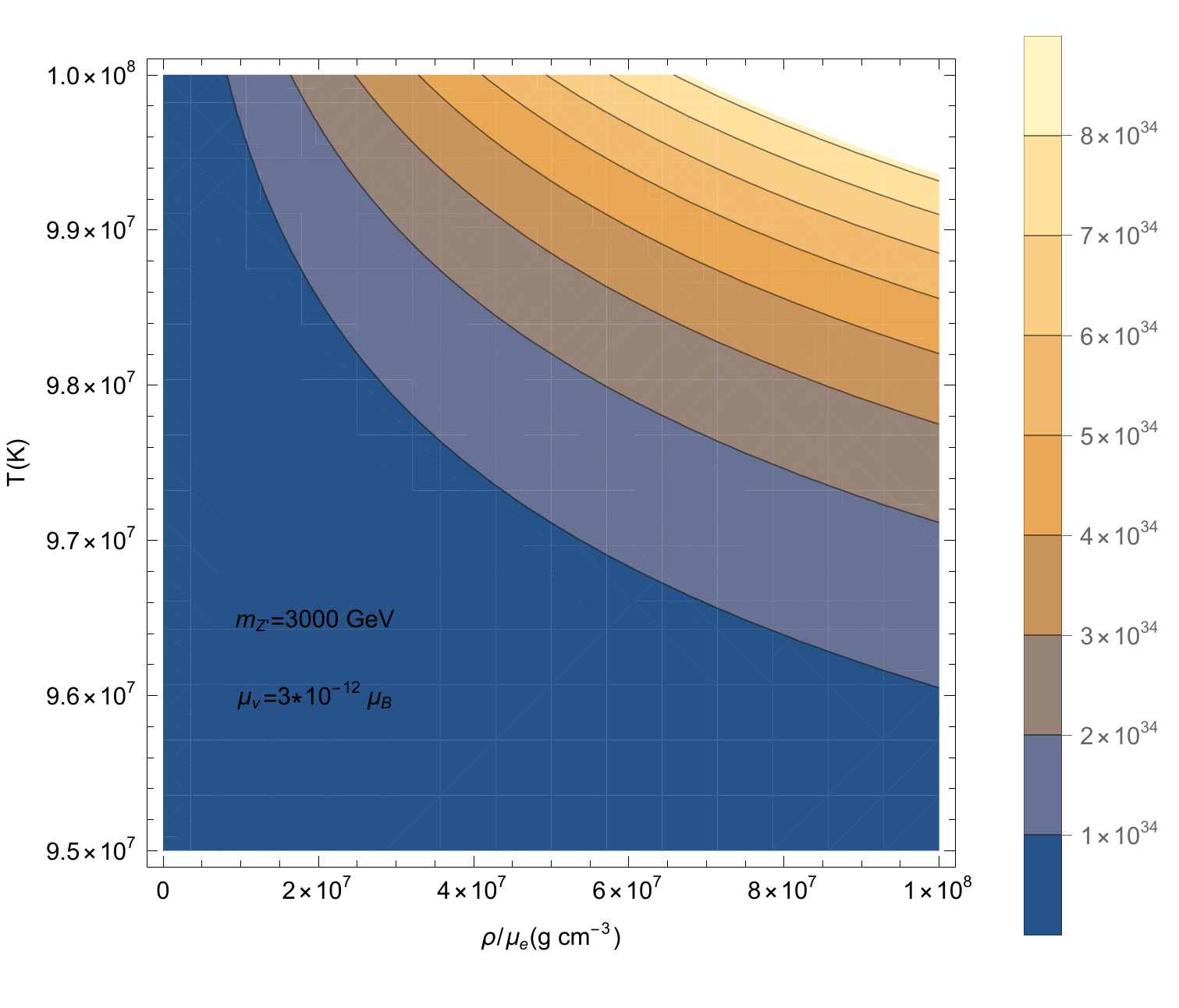}
\caption{Energy loss rate for region II as a function of temperature and density }
\label{QII331rhoTK}
\end{figure}

Next we plot the dependence of the energy loss on the temperature as in Fig. \ref{QITK}. We plot for three  values
 of the mass of the gauge boson $m_{Z'}=500, 100, 2000$ GeV. The ELR increases with the increase
 of the temperature. The dependence of $\mathcal{Q}$ on dimensionless parameter $\eta$ is depicted as in Fig. \ref{QII331eta}.
 As in the figure, $\mathcal{Q}$ decreases with the increase of $\eta$. This is what one would expect
  since $\eta=\fr{\mu_e}{k_B T} $ is inverse proportional to the temperature T. The dependence
  of $\mathcal{Q}$ on the temperature and density %is investigated
  are plotted  in Fig.\ref{QII331rhoTK}.
  $\mathcal{Q}$ increases in the region of high temperature and high density.

\section{Conclusions}
\label{conclusion}

Quantifying stellar loss energy is a priority in astrophysics and cosmology. One of most interesting
possibilities is to use stars and its physical process to put set constraints on  physics beyond Standard Model.

We have evaluated the stellar ELR in the frameworks  of the 3-3-1$\beta$ model. The energy loss
rate is in the form of neutrino emission assessed in the pair annihilation  $e^+  e^-\xrightarrow{\ga ,W,Z, Z'} \nu \bar{ \nu}$.
 We obtained the approximated formula for energy loss ($\mathcal{Q}$) and the correction
 $\mathcal{Q}$ in comparing with that of the SM.

We evaluated the $\mathcal{Q}$ correction for different value of $\beta=\pm\fr{1}{\sqrt{3}},\pm\fr{2}{\sqrt{3}},
\pm\sqrt{3}$. The negative value of $\beta$ give higher value compared with positive value and up to 14\%.
We have shown that the contribution of dipole moment is small compared with that of the  $Z'$ boson.
The $\mathcal{Q}$ gives the constraints on the mass range of the  $Z'$ boson $m_{Z'}\leq 4000$ GeV
which is in agreement with current searching the mass range of the $Z'$ at LHC.

\section*{ Acknowledgments}

We  acknowledge the financial support of the International Centre of Physics at the Institute of Physics, VAST under grant No: ICP.2021.02

\end{document}